\documentclass[graybox]{svmult}

\usepackage{mathptmx}       
\usepackage{helvet}         
\usepackage{courier}        
\usepackage{type1cm}        
\usepackage{makeidx}         
\usepackage{graphicx}        
\usepackage{multicol}        
\usepackage[bottom]{footmisc}

\def\be{\begin{equation}} 
\def\ee{\end{equation}} 
\def\bea{\begin{eqnarray}} 
\def\eea{\end{eqnarray}} 

\makeindex             

\begin{document}

\title*{Unconventional Cosmology}
\author{Robert H. Brandenberger}
\institute{Robert H. Brandenberger 
\at Physics Department, McGill University, 3600 University Street, Montreal, QC H3A 2T8, Canada \email{rhb@physics.mcgill.ca}}
%
%
\maketitle

\abstract*{I review two cosmological paradigms which are alternative
to the current inflationary scenario. The first alternative
is the ``matter bounce'', a non-singular bouncing cosmology
with a matter-dominated phase of contraction. The second
is an ``emergent'' scenario, which can be implemented in
the context of ``string gas cosmology''. I will compare
these scenarios with the inflationary one and demonstrate
that all three lead to an approximately scale-invariant spectrum
of cosmological perturbations.}

\abstract{I review two cosmological paradigms which are alternative
to the current inflationary scenario. The first alternative
is the ``matter bounce'', a non-singular bouncing cosmology
with a matter-dominated phase of contraction. The second
is an ``emergent'' scenario, which can be implemented in
the context of ``string gas cosmology''. I will compare
these scenarios with the inflationary one and demonstrate
that all three lead to an approximately scale-invariant spectrum
of cosmological perturbations.}

\section{Introduction}

\subsection{Overview}

``Unconventional cosmology" is the title which I was given for my
lectures. Based on my interpretation of this title my job is to
lecture on alternatives to the current paradigm of early universe
cosmology, the ``conventional theory". The fact that almost all
cosmologists agree that there is a current paradigm speaks to
the remarkable progress of cosmology over the past three decades.
At the time when the current paradigm of early universe cosmology,
the inflationary scenario \cite{Guth} (see also \cite{Brout, Starob1, Sato}),
was developed, we had very little observational information about
the large-scale structure of the universe. The success of inflationary
cosmology at that point in time is that it could explain some of the 
puzzles which the previous  paradigm - Standard Big Bang cosmology - 
could not address.  It was very soon realized
\cite{Mukh} (see also \cite{Press,Sato,Starob2})) that inflation
was much more powerful than simply being able to explain
puzzles of Standard Big Bang cosmology such as the flatness
and horizon problems. In fact, inflationary cosmology gave rise to 
the first explanation for the origin of inhomogeneities in the universe 
based on causal physics: It yields a mechanism for generating
an approximately scale-invariant spectrum of primordial
density fluctuations, i.e. the kind of spectrum which had already
been suggested as a reasonable one to be consistent with the
(at that time limited) information about the distribution of galaxies
\cite{Zel, Harrison}. As already realized earlier, such a
primordial spectrum of density fluctuations leads to an 
angular power spectrum of anisotropies in the cosmic microwave 
background (CMB) radiation which is scale-invariant on large scales 
and characterized by acoustic oscillations on angular scales
of a degree and lower \cite{SZ, Peebles}. This prediction has
has now been confirmed observationally \cite{WMAP} with high
accuracy. It is important, however, to keep in mind that any
theory which yields an approximately scale-invariant spectrum of
primordial fluctuations - and I will present a couple of such
theories in these lectures - will agree with the recent high-precision
observations of the large-scale structure and CMB anisotropies.
Thus, current observations cannot be interpreted as a
proof that inflation took place.

In spite of its phenomenological success, inflationary cosmology
suffers from some important conceptual problems, which may
imply that it is not so ``conventional" after all. These problems
motivate the search for alternative proposals for the evolution
of the early universe and for the generation of structure. These
alternatives must be consistent with the current observations,
and they must make predictions with which they can be
observationally distinguished from inflationary cosmology.

There are indeed paradigms alternative to inflation which generate
an almost scale-invariant spectrum of primordial cosmological
fluctuations. In these lectures I will present two examples -
first the string gas realization \cite{BV, NBV} (see \cite{SGCrev}
for reviews) of the emergent universe paradigm \cite{emergent}, and second the 
``matter bounce scenario" \cite{Wands, Fabio1} (see \cite{ON}
for reviews). I should emphasize, however, that these are not the
only alternatives to the inflationary scenario. The Pre-Big-Bang
scenario \cite{PBB}, the Ekpyrotic scenario \cite{Ekp}, and the pseudo-conformal
construction \cite{Rubakov} are other promising models, and
there are others.

The outline of this lecture series is as follows. The first lecture
(Sections 1 - 3)
focuses on background (homogeneous and isotropic) cosmologies.
I begin with a review of the inflationary scenario, the current paradigm
of early universe cosmology. After discussing the phenomenological
successes of the scenario, I will list a number of conceptual problems
which in part motivate the search for alternative scenarios. In Section 2
I introduce the first alternative paradigm which will be discussed here,
the ``matter bounce" scenario. After presenting the basic idea of the
scenario, I will discuss various ways to realize it. In Section 3 I turn
to the ``emergent Universe" scenario. Once again, I begin by presenting
the basic ideas before turning to a discussion of ``string gas cosmology",
the specific realization which has provided some very interesting results.

The second lecture (Sections 4 - 7)
focuses on the question of how the inhomogeneities
and anisotropies which are observed now in the distribution of
galaxies on large scales and in temperature maps of the CMB, respectively,
are generated. I will first (Section 4) briefly review the theory of
cosmological perturbations. Then, I will emphasize that all three scenarios (inflation,
the matter bounce and string gas cosmology), yield fluctuations in
agreement with current data, but are distinguishable by future
observations. Fluctuations in inflation are reviewed in Section 5, those in the 
matter bounce in Section 6, and those in string gas cosmology in Section 7. 
The final section focuses on outstanding problems of the
various scenarios, and contains some general discussion.

These lectures are a modified version of lectures given previously
\cite{ON} at various summer schools. 

\subsection{Review of Inflationary Cosmology}

Inflationary cosmology \cite{Guth} addresses several shortcomings
of Standard Big Bang cosmology (the previous paradigm of
early universe cosmology). It explains why the universe is
to a good approximation homogeneous and isotropic on large
scales (the ``horizon problem"), 
it explains why it is to an excellent accuracy spatially
flat (the ``flatness problem"), and it can explain its large size and 
entropy from initial conditions where the universe is of microscopic size.
 
The idea of inflationary cosmology is to add a period to the evolution of
the very early universe during which the scale factor undergoes
accelerated expansion - most often nearly exponential growth. 
To obtain exponential expansion of space in the context
of Einstein gravity, the energy density must be constant.
Thus, during inflation the total energy and size of the
universe both increase exponentially. In this way,
inflation can solve the size and entropy problems
of Standard Cosmology. Since the horizon expands
exponentially during the period of inflation and all classical
fluctuations redshift, inflation produces
an approximately homogeneous and isotropic space. In
addition, the relative contribution of spatial curvature decreases
during the period of inflation. Thus, inflation can also address
the ``flatness problem" of Standard Big Bang cosmology.
Any ``unconventional cosmology" which claims to provide
an alternative to inflation must also address the basic
problems of Standard Cosmology mentioned above.

\begin{figure}
\includegraphics[height=4cm, width=12cm]{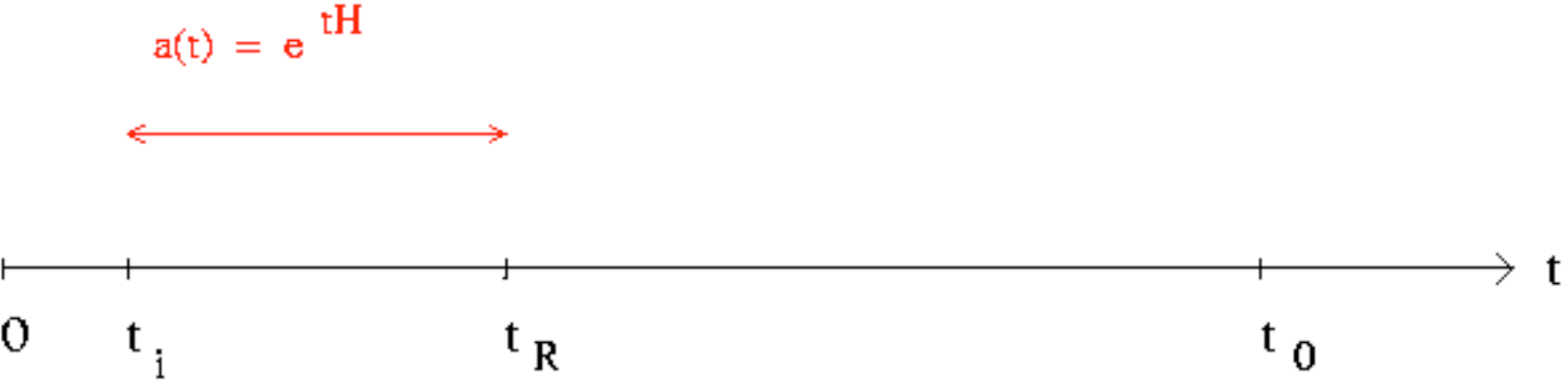}
\caption{A sketch showing the time line of inflationary cosmology.
The period of accelerated expansion begins at time $t_i$ and
end at $t_R$. The time evolution after $t_R$ corresponds to
what happens in Standard Cosmology.}
\label{timeline}
\end{figure}

The time line of inflationary cosmology is sketched in Figure \ref{timeline}.
The time $t_i$ is the beginning of the inflationary period, and
$t_R$ is its end (the meaning of the subscript $R$ will become
clear later). Although inflation is usually associated with physics
at very high energy scales, e.g. $E \sim 10^{16} {\rm Gev}$, all that
is required from the initial basic considerations is that inflation
ends before the time of nucleosynthesis.

During the period of inflation, the density of any pre-existing particles is
diluted exponentially. Hence, if inflation is to be viable, it must contain a
mechanism to heat the universe at $t_R$, a ``reheating"
mechanism - hence the subscript $R$ on $t_R$. This mechanism must
involve dramatic entropy generation. It is this non-adiabatic evolution
which leads to a solution of the flatness problem.

Inflationary cosmology, however, does more than simply solve
some conceptual problems of the previous paradigm. It for
the first time provided a causal theory of structure formation.
Any proposed alternative to cosmological inflation must also
match this success. Here we review the basic idea of
why inflationary cosmology can provide a causal explanation
of the observed inhomogeneities in the universe. The
calculations will be reviewed in the second lecture.

\begin{figure} 
\includegraphics[height=14cm]{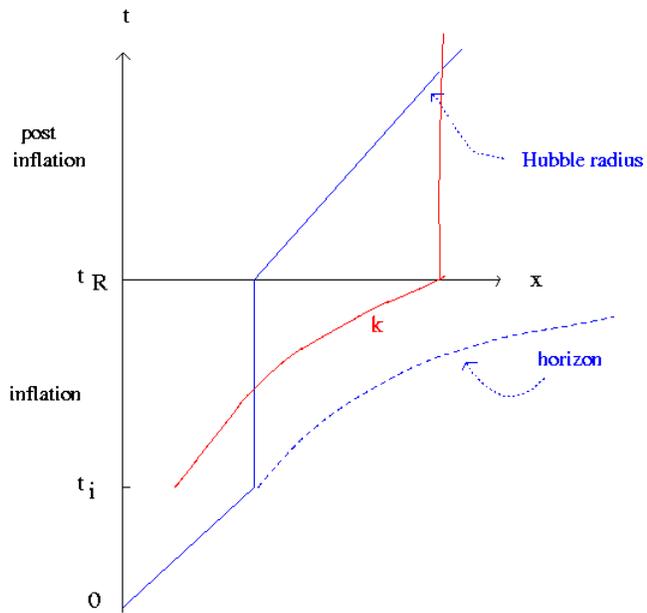}
\caption{Space-time sketch of inflationary cosmology.
The vertical axis is time, the horizontal axis corresponds
to physical distance. The solid line labelled $k$ is the
physical length of a fixed comoving fluctuation scale. The
role of the Hubble radius and the horizon are discussed
in the text.}
\label{infl1}
\end{figure}

In order to understand why inflation provides a causal
structure formation mechanism, we start with a
space-time sketch of inflationary cosmology as presented in
Figure \ref{infl1} . The vertical axis is time, the horizontal
axis corresponds to physical distance.  Three different
distance scales are shown. The solid line labelled by $k$
is the physical length corresponding to a fixed comoving
perturbation. The second solid line (blue) is the Hubble
radius 
\be
l_H(t) \, \equiv \, H^{-1}(t) \, .
\ee
As will be shown in Lecture 2, he Hubble radius separates 
scales where microphysics
dominates over gravity (sub-Hubble scales) from ones 
on which the effects of microphysics are negligible 
(super-Hubble scales). Hence, a
necessary requirement for a causal theory of structure
formation is that scales we observe today originate
at sub-Hubble lengths in the early universe. The third
length is the ``horizon", the forward light cone of our position
at the Big Bang. The horizon is the causality limit. Note
that because of the exponential expansion of space during
inflation, the horizon is exponentially larger than the
Hubble radius. It is important not to confuse these two
scales. Hubble radius and horizon are the same in
Standard Cosmology, but in all three early universe
scenarios which will be discussed in these lectures they
are completely different (in inflationary cosmology
the horizon is exponentially larger, in the matter bounce
scenario it is in fact infinite, and in the emergent scenario
it is infinite if the emergent phase extends to $ t = - \infty$).
In fact, in any structure formation scenario the two
scales need to be different.

{F}rom Fig. \ref{infl1} it is clear that provided the period
of inflation is sufficiently long, all scales which are currently
observed originate as sub-Hubble scales at the beginning
of the inflationary phase. Thus, in inflationary cosmology 
it is possible to have a causal generation mechanism of
fluctuations \cite{Mukh,Press,Sato}. Since matter pre-existing
at $t_i$ is redshifted away, we are left with a matter vacuum.
The inflationary universe scenario of structure formation is
based on the hypothesis that all current structure originated
as quantum vacuum fluctuations. From Figure \ref{infl1} it
is also clear that the horizon problem of standard cosmology
can be solved provided that the period of inflation lasts
sufficiently long. For inflation to solve the horizon and
flatness problem of Standard cosmology,
the period of exponential expansion must be greater than
about $50 H^{-1}$ (this number depends very slightly
on the energy scale at which inflation takes place).

In order to obtain exponential expansion of space
in the context of Einstein gravity, matter with an
equation of state $p = - \rho$ is required, where
$p$ and $\rho$ are pressure and energy density,
respectively. In the context of renormalizable quantum
field theory, a phase dominated by almost constant
(both in space and time) potential energy of a scalar matter field 
is required.  

\subsection{Conceptual Problems of Inflationary Cosmology}

In spite of the phenomenological success of the
inflationary paradigm, conventional scalar field-driven
inflation suffers from several important conceptual problems.

The first problem concern the nature of the inflaton, the scalar
field which generates the inflationary expansion. No particle
corresponding to the excitation of a scalar field has yet been
observed in nature, and the Higgs field which is introduced
to give elementary particles masses in the Standard Model of
particle physics does not have the required flatness of the
potential to yield inflation, unless it is non-minimally coupled
to gravity \cite{Shaposh}. In particle physics theories beyond
the Standard Model there are often many scalar fields,
but it is in general very hard to obtain the required flatness
properties on the potential 

The second problem (the {\bf amplitude problem})
relates to the amplitude of the spectrum of
cosmological perturbations. In a wide class of inflationary
models, obtaining the correct amplitude requires the introduction
of a hierarchy in scales, namely \cite{Adams}
\be
{{V(\varphi)} \over {\Delta \varphi^4}}
\, \leq \, 10^{-12} \, ,
\ee
where $\Delta \varphi$ is the change in the inflaton field during
the minimal length of the inflationary period, and $V(\varphi)$ is
the potential energy during inflation.

\begin{figure}
\includegraphics[height=14cm]{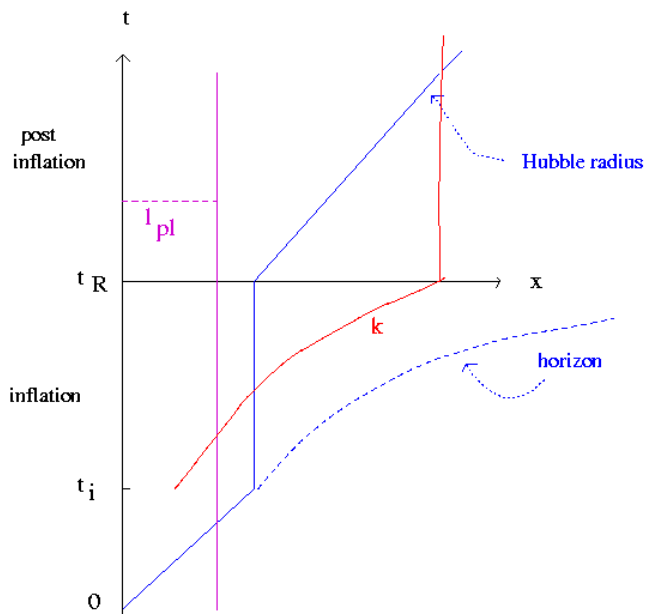}
\caption{Space-time diagram (sketch) 
of inflationary cosmology where we have added an extra
length scale, namely the Planck length $l_{pl}$ (majenta vertical line).
The symbols have the same meaning as in Figure 2.
Note, specifically, that - as long as the period of inflation
lasts a couple of e-foldings longer than the minimal value
required for inflation to address the problems of Standard
Big Bang cosmology - all wavelengths of cosmological interest
to us today start out at the beginning of the period of inflation
with a wavelength which is smaller than the Planck length.}
\label{infl2}       
\end{figure}

A more serious problem is the {\bf trans-Planckian problem} \cite{RHBrev3}.
Returning to the space-time diagram of inflation (see Figure \ref{infl2}), 
we can immediately
deduce that, provided that the period of inflation lasted sufficiently
long (for GUT scale inflation the number is about 70 e-foldings),
then all scales inside the Hubble radius today started out with a
physical wavelength smaller than the Planck scale at the beginning of
inflation. Now, the theory of cosmological perturbations is based
on Einstein's theory of General Relativity coupled to a simple
semi-classical description of matter. It is clear that these
building blocks of the theory are inapplicable on scales comparable
and smaller than the Planck scale. Thus, the key
successful prediction of inflation (the theory of the origin of
fluctuations) is based on suspect calculations since 
new physics {\it must} enter
into a correct computation of the spectrum of cosmological perturbations.
The key question is as to whether the predictions obtained using
the current theory are sensitive to the specifics of the unknown
theory which takes over on small scales. Simple toy models
of new physics on super-Planck scales based on modified
dispersion relations were used in \cite{Jerome1} (see also
\cite{Niemeyer}) to show that the resulting spectrum of
cosmological fluctuations indeed depends on what is assumed
about physics on trans-Planckian scales.
 
A fourth problem is the {\bf singularity problem}. It was known for a long
time that Standard Big Bang cosmology cannot be the complete story of
the early universe because of
the initial singularity, a singularity which is unavoidable when basing
cosmology on Einstein's field equations in the presence of a matter
source obeying the weak energy conditions (see e.g. \cite{HE} for
a textbook discussion). The singularity theorems have been
generalized to apply to Einstein gravity coupled to scalar field
matter, i.e. to scalar field-driven inflationary cosmology \cite{Borde}.
It was shown that, in this context, a past singularity at some point
in space is unavoidable. Thus we know, from the outset, that scalar
field-driven inflation cannot be the ultimate theory of the very
early universe.

The Achilles heel of scalar field-driven inflationary cosmology may
be the {\bf cosmological constant problem}. We know from
observations that the large quantum vacuum energy of field theories
does not gravitate today. However, to obtain a period of inflation
one is using the part of the energy-momentum tensor of the scalar field
which looks like the vacuum energy. In the absence of a 
solution of the cosmological constant problem it is unclear whether
scalar field-driven inflation is robust, i.e. whether the
mechanism which renders the quantum vacuum energy gravitationally
inert today will not also prevent the vacuum energy from
gravitating during the period of slow-rolling of the inflaton 
field.

A final problem which we will mention here is the concern that the
energy scale at which inflation takes place is too high to justify
an effective field theory analysis based on Einstein gravity. In
simple toy models of inflation, the energy scale during the period
of inflation is about $10^{16} \rm{GeV}$, very close to the string scale
in many string models, and not too far from the Planck scale. Thus,
correction terms in the effective action for matter and gravity may 
already be important at the energy scale of inflation, and the 
cosmological dynamics may be rather different from what is
obtained when neglecting the correction terms.

\begin{figure}
\includegraphics[height=14cm]{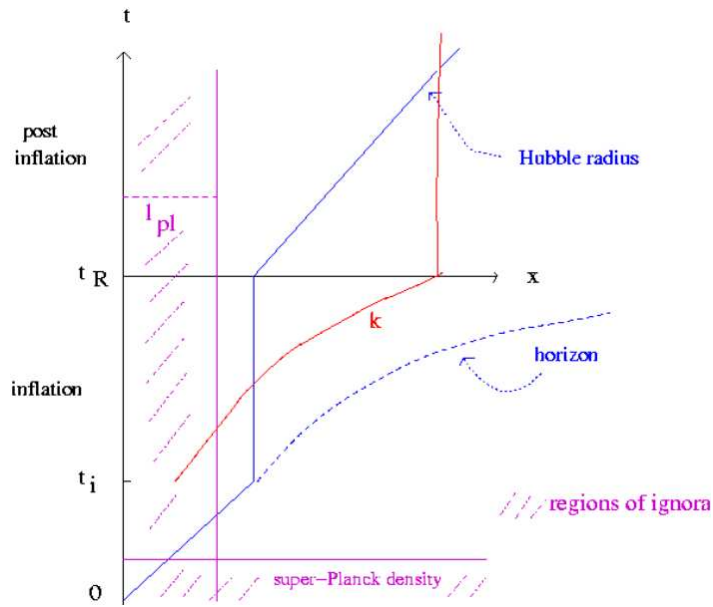}
\caption{Space-time diagram (sketch) 
of inflationary cosmology including the two zones of
ignorance - sub-Planckian wavelengths and trans-Planckian 
densities. The symbols have the same meaning as in Figure 2.
Note, specifically, that - as long as the period of inflation
lasts a couple of e-foldings longer than the minimal value
required for inflation to address the problems of Standard
Big Bang cosmology - all wavelengths of cosmological interest
to us today start out at the beginning of the period of inflation
with a wavelength which is in the zone of ignorance.}
\label{infl3}       
\end{figure}

In Figure \ref{infl3} we show once again the space-time sketch
of inflationary cosmology. In addition to the length scales
which appear in the previous versions of this figure,
we have now shaded the ``zones of ignorance", zones
where the Einstein gravity effective action is sure to
break down. As described above, fluctuations emerge
from the short distance zone of ignorance (except if
the period of inflation is very short), and the energy
scale of inflation might put the period of inflation too
close to the high energy density zone of ignorance to
trust the predictions based on using the Einstein
action. 

The arguments in this subsection provide a motivation
for considering alternative scenarios of early universe
cosmology. Below we will focus on two scenarios, the
{\it matter bounce} and {\it string gas cosmology},
a realization of the {\it emergent universe} paradigm.

\section{Matter Bounce}

\subsection{The Idea}

The first alternative to cosmological
inflation as a theory of structure formation is the
 ``matter bounce" , an alternative which is not
yet well appreciated (for an overview the reader is
referred to \cite{ON}). The scenario is
based on a cosmological background in which the
scale factor $a(t)$ bounces in a non-singular manner.

\begin{figure}[htbp] 
\includegraphics[height=9cm]{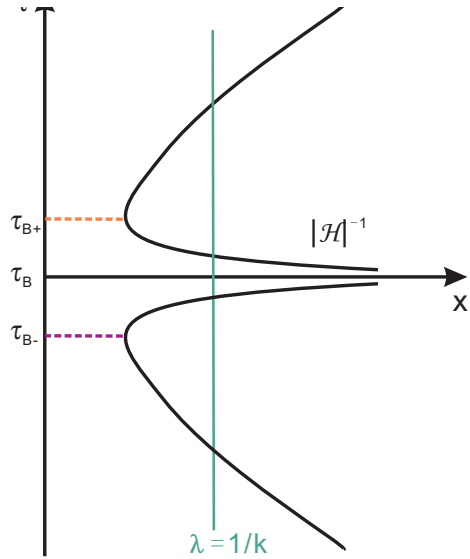}
\caption{Space-time sketch in the matter bounce scenario. The vertical axis
is conformal time $\eta$, the horizontal axis denotes a co-moving space coordinate.
The vertical line indicates the wavelength of a fluctuation mode.
Also, ${\cal H}^{-1}$ denotes the co-moving Hubble radius.}
\label{bounce}
\end{figure}

{F}igure \ref{bounce} shows a space-time sketch
of such a bouncing cosmology. Without loss of generality
we can adjust the time axis such that the bounce point
(minimal value of the scale factor) occurs at $t = 0$.
There are three phases in such a non-singular bounce: the initial
contracting phase during which the Hubble radius is
decreasing linearly in $|t|$, a bounce phase during which
a transition from contraction to expansion takes place,
and thirdly the usual expanding phase of Standard
Cosmology. There is no prolonged inflationary phase 
after the bounce, nor is there a time-symmetric deflationary 
contracting period before the bounce point. As is obvious
from the Figure, scales which we observe today started
out early in the contracting phase at sub-Hubble lengths.
The matter bounce scenario assumes that the contracting
phase is matter-dominated at the times when scales
we observe today exit the Hubble radius. A model in
which the contracting phase is the time reverse of our
current expanding phase would obey this condition.
The assumption of an initial matter-dominated phase
will be seen later in Lecture 2 to be important if
we want to obtain a scale-invariant spectrum of
cosmological perturbations from initial vacuum
fluctuations \cite{Wands,Fabio1}.

Let us make a first comparison with inflation. A non-deflationary
contracting phase replaces the accelerated expanding phase
as a mechanism to bring fixed comoving scales within the Hubble
radius as we go back in time, allowing us to consider the possibility
of a causal generation mechanism of fluctuations. Starting with
vacuum fluctuations, a matter-dominated contracting phase is
required in order to obtain a scale-invariant spectrum. This corresponds
to the requirement in inflationary cosmology that the accelerated expansion be
nearly exponential.

How are the problems of Standard Big Bang cosmology addressed in
the matter bounce scenario? First of all, note that since
the universe begins cold and large, the size and entropy problems 
of Standard Cosmology do not arise.
As is obvious from Figure \ref{bounce}, there is no horizon problem
for the matter bounce scenario as long as the contracting period is long
(to be specific, of similar duration as the post-bounce expanding
phase until the present time). By the same argument, it is possible
to have a causal mechanism for generating the primordial
cosmological perturbations which evolve into the structures we observe
today. Specifically, as will be discussed in Section 6,
if the fluctuations originate as vacuum perturbations
on sub-Hubble scales in the contracting phase, then the
resulting spectrum at late times for scales exiting the Hubble
radius in the matter-dominated phase of contraction is
scale-invariant \cite{Wands,Fabio1}.  

The flatness problem is
the one which is only partially addressed in the matter bounce
setup. The contribution of the spatial curvature decreases
in the contracting phase at the same rate as it increases in
the expanding phase. Thus, to explain the observed spatial
flatness, comparable spatial flatness at early times in the
contracting phase is required. This is an improved situation compared
to the situation in Standard Big Bang cosmology where spatial
flatness is overall an unstable fixed point and hence extreme
fine tuning of the initial conditions is required to explain
the observed degree of flatness. But the situation is not as
good as it is in a model with a long period of inflation where spatial
flatness is a local attractor in initial condition space (it is not
a global attractor, though!).

How does the matter bounce scenario address the conceptual
problem of inflation? First of all, the length scale of
fluctuations of interest for current observations on cosmological
scales is many orders of magnitude larger than the Planck
length throughout the evolution. If the energy scale at the
bounce point is comparable to the particle physics GUT
scale, then typical wavelengths at the bounce point are
not too different from $1 {\rm mm}$. Hence, the fluctuations
never get close to the small wavelength zone of ignorance in
Figures 3 and 4, and thus a description of the evolution of
fluctuations using Einstein gravity should be well justified
modulo possible difficulties at the bounce point which we will
return to in Section 6. Thus, there is no trans-Planckian problem
for fluctuations in the matter bounce scenario.

As will be discussed below, new physics is required in order to
provide a non-singular bounce. Thus, the ``solution" of the
singularity problem is put in by hand and cannot be counted
as a success, except in realizations of the matter bounce in
the context of a string theory background in which the non-singular
evolution follows from general principles. Such a 
theory has recently been presented in \cite{KPT} (see
\cite{BKPPT} for an analysis of fluctuations in these models).
Existing matter bounce models do not address the cosmological
constant problem. However, I would like to emphasize that
the mechanism which drives the evolution in the matter bounce
scenario is robust against our ignorance of what solves the
cosmological constant problem, an improvement of the
situation compared to the situation in inflationary cosmology.

With Einstein gravity and matter satisfying the usual energy conditions
it is not possible to obtain a non-singular bounce. Thus, new
physics is required in order to obtain a non-singular bouncing
cosmology. Such new physics can arise by modifying the 
gravitational sector, or by modifying the matter sector.
The study of bouncing cosmologies has a long history
(see \cite{Novello} for an in-depth review of a lot of these
past approaches). We will now turn to a brief overview of
some more recent work on non-singular bouncing cosmology
with the matter bounce in mind.

\subsection{Realizing a Matter Bounce with Modified Matter}

In order to obtain a cosmological bounce in the context of
Einstein gravity, it is necessary to introduce a new form of
matter which violates the Null Energy Condition (NEC). 
A simple way to do this is by introducing quintom matter \cite{quintom}. 
Resulting nonsingular quintom bouncing models have been
discussed in \cite{Cai1}. Quintom matter is a set of two matter
fields, one of them regular matter (obeying the NEC), 
the second a ``phantom'' field with opposite sign kinetic
term which violates the NEC. Even though this
model is plagued by ghost instabilities \cite{ghost},
we will use it to illustrate the basic idea of how a
bouncing cosmology can be obtained.

We \cite{Cai1} model both matter components with
scalar fields, the mass of the regular one 
($\varphi$) being $m$, 
and $M$ being that of the field ${\tilde{\varphi}}$ with 
wrong sign kinetic term. We consider a contracting universe
and assume that early on both fields are oscillating, but that the amplitude $\cal{A}$
of $\varphi$ greatly exceeds the corresponding amplitude
${\tilde{\cal A}}$ of ${\tilde{\varphi}}$ such that the energy density
is dominated by $\varphi$. During the initial period
of contraction, both amplitudes grow at the
same rate. At some point, $\cal{A}$ will become so large 
that the oscillations of $\varphi$ freeze out \footnote{This
corresponds to the time reverse of
entering a region of large-field inflation.}.
Then, $\cal{A}$ will grow only slowly, whereas ${\tilde{\cal A}}$
will continue to increase. Thus, the (negative) energy density
in ${\tilde{\varphi}}$ will grow in absolute value relative to that
of $\varphi$. The total energy density will decrease towards $0$.
At that point, $H = 0$ by the Friedmann equations. Since
it is only the phantom field which has large kinetic energy,
it follows that ${\dot{H}} > 0$ when $H = 0$. Hence,
a non-singular bounce occurs. 

The Higgs sector of the
Lee-Wick model \cite{LW} provides a concrete realization of 
the quintom bounce model, as studied in \cite{LWbounce}.
Quintom models like all other models with negative sign kinetic
terms suffer from an instability problem \cite{ghost} in the
matter sector and are hence problematic. In addition, they are
unstable against the addition of radiation (see e.g. \cite{Karouby})
and anisotropic stress (the BKL instability \cite{BKL}).

An improved way of obtaining a non-singular bouncing cosmology
using modified matter is by using a ghost condensate field \cite{Chunshan}
(see also \cite{Senatore, NewEkp} where ghost condensates have been
used to produce non-singular bounces in different contexts).
The ghost condensation mechanism is the analog of the Higgs
mechanism in the kinetic sector of the theory. In the Higgs
mechanism we take a field $\phi$ whose mass when evaluated
at $\phi = 0$ is tachyonic, add higher powers of $\phi^2$ to the
potential term in the Lagrangian such that there is a stable
fixed point $\phi = v \neq 0$, and thus when expanded about
$\phi = v$ the mass term has the ``safe" non-tachyonic sign.
In the ghost condensate construction we take a field $\phi$
whose kinetic term
\be
X \, \equiv \,  - g^{\mu \nu} \partial_{\mu} \phi \partial_{\nu} \phi 
\ee
appears with the wrong sign in the Lagrangian. Then, we add
higher powers of $X$ to the kinetic Lagrangian such that there
is a stable fixed point $X = c^2$ and such that when expanded
about $X = c^2$ the fluctuations have the regular sign of the kinetic
term:
\be
{\cal L} \, = \, \frac{1}{8} M^4 \bigl( X - c^2 \bigr)^2 - V(\phi) \, ,
\ee
where $V(\phi)$ is a usual potential function, $M$ is a 
characteristic mass scale and the dimensions of $\phi$
are chosen such that $X$ is dimensionless.

In the context of cosmology, the ghost condensate is
\be
\phi \, = \, c t 
\ee
and breaks local Lorentz invariance. Now let us expand the
homogeneous component of $\phi$ about the ghost condensate:
\be
\phi(t) \, = \, c t + \pi(t) \, .
\ee
If ${\dot \pi} < 0$ then the gravitational energy density is negative,
and a non-singular bounce is possible. Thus, in \cite{Chunshan}
we constructed a model in which the ghost condensate field
starts at negative values and the potential $V(\phi)$ is
negligible. As $\phi$ approaches $\phi = 0$ it encounters a
positive potential which slows it down, leading to ${\dot \pi} < 0$
and hence to negative gravitational energy density. Thus,
a non-singular bounce can occur. We take the potential
to be of the form
\be
V(\phi) \, \sim \, \phi^{- \alpha}
\ee
 for $|\phi| \gg M$, where $M$ is the mass scale above which
 the higher derivative kinetic terms are important. For
 sufficiently large values of $\alpha$, namely
 \be
 \alpha \, \geq \, 6 \, ,
 \ee
 the energy density in the ghost condensate increases faster
 than that of radiation and anisotropic stress at the universe
 contracts . Hence, this bouncing cosmology is locally stable
 against the addition of radiation and anisotropic stress (there
is still an instability to the development of
anisotropic stress in the contracting phase prior to the
time when the ghost condensate starts to dominate).

Non-singular bouncing cosmologies can also be obtained
making use of Galileon models \cite{Galileonbounce}.
However, these models also suffer from an instability against 
the development of anisotropic stress.

The Ekpyrotic contracting universe (contracting phase with an
equation of state $w \gg 1$ is stable against the growth
of anisotropies, as shown in \cite{EkpStable}). Thus, one
way of obtaining a matter bounce which is stable against
the development of anisotropic stress is to have a phase
of Ekpyrotic contraction set in shortly after the time $t^{-}_{eq}$
of equal matter and radiation in the contracting phase. A
model in which this is realized and in which the non-singular
bounce is generated by a ghost condensate and Galileon
construction has recently been worked out in \cite{Yifu4}.

\subsection{Realizing a Matter Bounce with Modified Gravity}

It is unreasonable to expect that Einstein gravity
will provide a good description of the physics at very
high energy densities. In particular, all approaches to
quantum gravity lead to correction terms in the
gravitational action (compared to the pure Einstein term)
which become dominant at the Planck scale. 
It is possible (and in some approaches to quantum
gravity such as string theory even likely) that the
new terms will tend to prevent cosmological
singularities from appearing, and hence might
allow a bouncing cosmology even in the presence
of matter which obeys the NEC. 

One early study is based on a higher derivative Lagrangian 
resulting from the ``nonsingular universe construction" of \cite{MBS}
which is based on a Lagrange multiplier construction which forces
all space-time curvature invariants to stay bounded as the
the energy density increases. This Lagrangian admits
bouncing solutions in the presence of regular matter. Another
model is the non-local higher derivative action of \cite{Biswas1}  
which is constructed to be ghost-free about Minkowski space-time
and which admits bouncing solutions.  Mirage cosmology
\cite{mirage} (induced gravity on a brane which is moving into 
and out of a high-curvature throat of a higher-dimensional 
bulk space-time also admits bouncing cosmologies \cite{Saremi}. 

A few years ago there was a lot of interest in 
Horava-Lifshitz gravity \cite{Horava}, an approach to
quantum gravity in four space-time dimensions which
is based on a gravitational Lagrangian which is power-counting
renormalizable with respect to the reduced symmetry group
of spatial diffeomorphisms only (we drop the invariance
requirement under space-dependent time reparametrizations),
The lost symmetry is replaced by an anisotropic
scaling symmetry between space and time. The Lagrangian contains
higher space derivative terms. As was realized in \cite{RHBHL},
in the presence of spatial curvature these higher space
derivative terms act as ghost radiation and ghost anisotropic
stress and lead to the possibility of a non-singular bouncing cosmology.

Loop quantum cosmology is an approach to quantum cosmology
which also leads to bouncing solutions (see e.g. 
\cite{LQCreview} for a review). What is responsible here for
singularity avoidance is the fundamental discreteness of
the area which comes from quantization. Other
lecturers at this school have discussed loop quantum cosmology
in depth.

Superstring theory as a quantum theory which includes gravity
will likely also resolve cosmological singularities. As will be
discussed in detail in the section on string gas cosmology,
the new degrees of freedom which string theory admits
compared to point particle theories lead to duality
symmetries which relate large and small spaces. Physical
quantities such as the temperature remain bounded,
and it is hence likely to obtain bouncing cosmological
solutions. Our understanding of string cosmology is
hampered by the lack of a fully non-perturbative
formulation of string theory in a cosmological
space-time. Most analyses of string cosmology
are performed using string-motivated field theory.
A specific theory in which the field theory approximations
are under good control is the Type II string cosmology
of \cite{KPT}.

\section{Emergent Universe}

\subsection{The Idea}

\begin{figure}
\includegraphics[height=6cm]{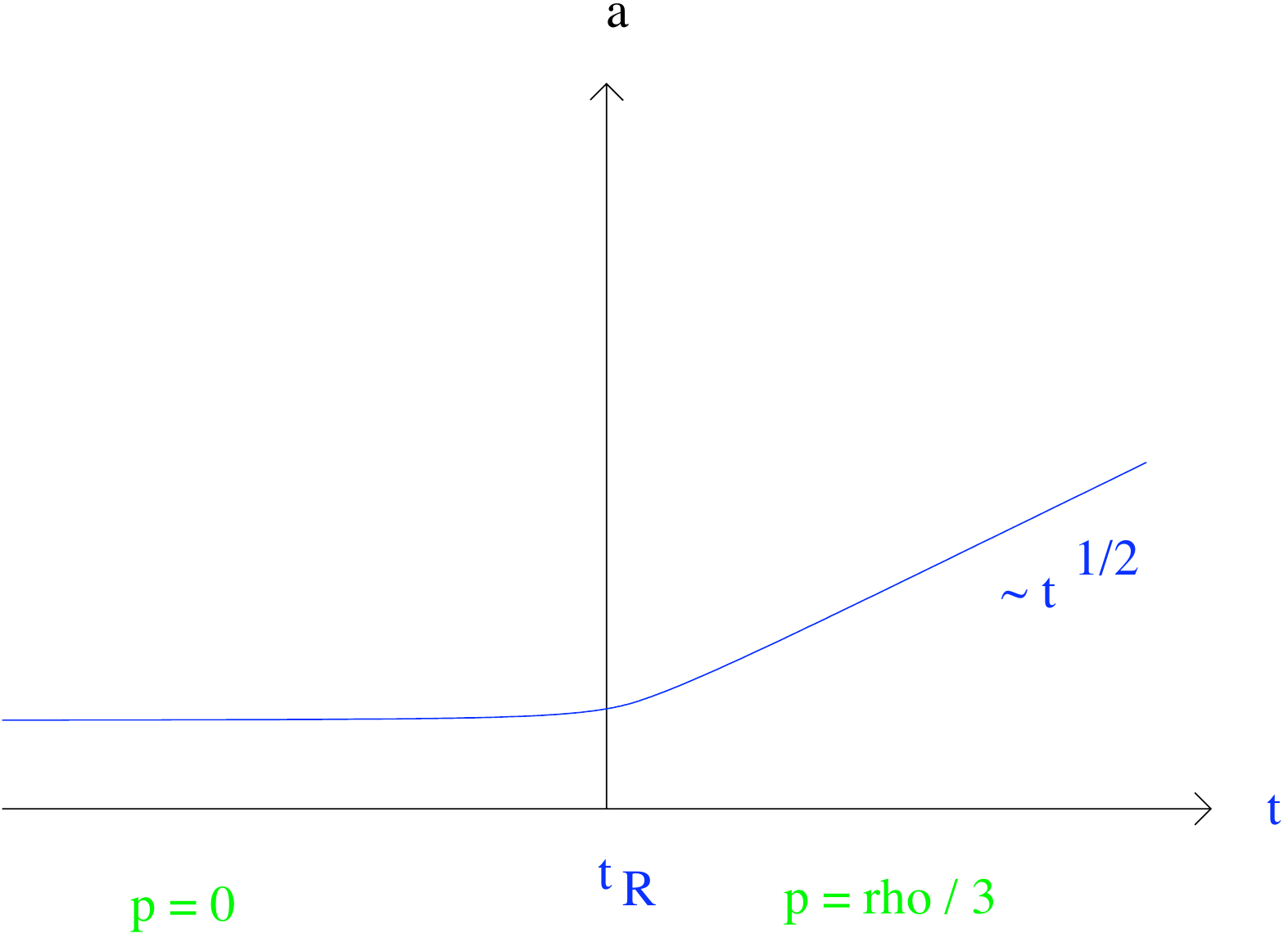}
\caption{The dynamics of emergent universe cosmology. The vertical axis
represents the scale factor of the universe, the horizontal axis is
time. }
 \label{timeevol}
\end{figure}

The ``emergent universe" scenario \cite{emergent} is another
non-singular cosmological scenario in which time runs
from $- \infty$ to $+ \infty$. The idea is that if we follow
the evolution of our homogeneous and isotropic space-time
into the past, the expansion rate $H$ ceases to increase as we
approach a certain limiting scale (most likely related to
the Planck energy). Instead of further increasing, $H$
decreases to zero, and the scale factor approaches a
constant value as we tend to past infinity.  The time
evolution of the scale factor is sketched in Figure (\ref{timeevol}).

\begin{figure} 
\includegraphics[height=10cm]{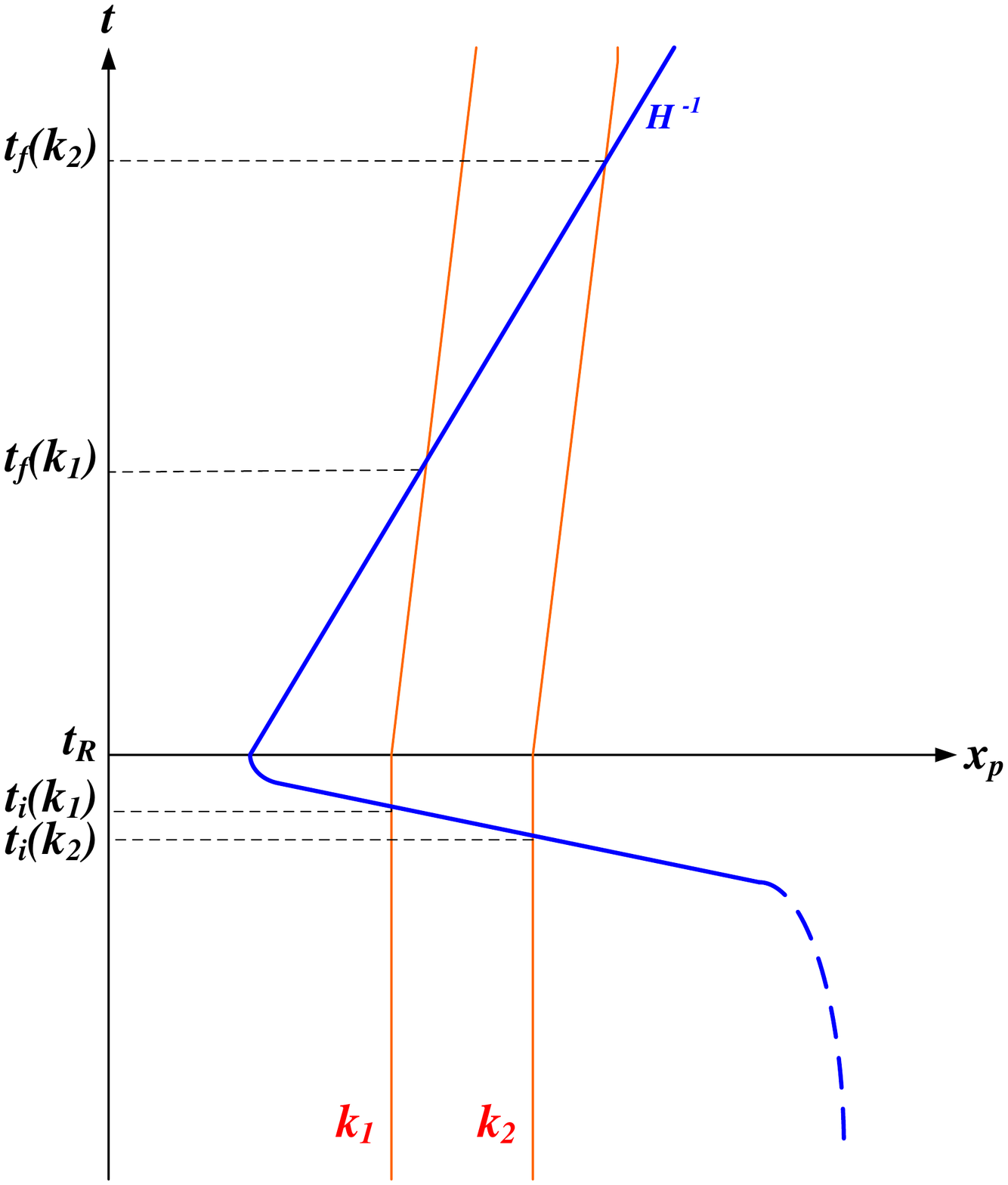}
\caption{Space-time diagram (sketch) showing the evolution of fixed 
co-moving scales in emergent cosmology. The vertical axis is time, 
the horizontal axis is physical distance.  
The solid curve represents the Einstein frame Hubble radius 
$H^{-1}$ which shrinks abruptly to a micro-physical scale at $t_R$ and then 
increases linearly in time for $t > t_R$. Fixed co-moving scales (the 
dotted lines labeled by $k_1$ and $k_2$) which are currently probed 
in cosmological observations have wavelengths which were smaller than 
the Hubble radius long before $t_R$. They exit the Hubble 
radius at times $t_i(k)$ just prior to $t_R$, and propagate with a 
wavelength larger than the Hubble radius until they re-enter the 
Hubble radius at times $t_f(k)$.}
\label{spacetimenew}
\end{figure}

In Figure \ref{spacetimenew} we sketch the space-time diagram
in emergent cosmology. Since the early emergent phase is
quasi-static, the Hubble radius is infinite. For the same reason,
the physical wavelength of fluctuations remains constant in
this phase. At the end of the emergent phase, the Hubble radius
decreases to a microscopic value and makes a transition to
its evolution in Standard Cosmology.

Once again, we see that fluctuations originate on sub-Hubble scales.
In emergent cosmology, it is the existence of a quasi-static phase
which leads to this result. What sources fluctuations depends on
the realization of the emergent scenario. String gas cosmology
is the example which I will consider later on. 
In this case, the source of perturbations 
is thermal: string thermodynamical fluctuations in
a compact space with stable winding modes, and this
in fact leads to a scale-invariant spectrum \cite{NBV}.

How does emergent cosmology address the problems of
Standard Cosmology? As in the case of a bouncing 
cosmology, the horizon is infinite and hence there is
no horizon problem. Since there is likely thermal
equilibrium in the emergent phase, a mechanism
to homogenize the universe exists, and hence spatial
flatness is not a mystery. As discussed in the previous
paragraph, there is no causality obstacle against
producing cosmological fluctuations. The scenario
is non-singular, but this cannot in general be weighted
as a success unless the emergent phase can be shown
to arise from some well controlled ultraviolet physics.

Like in the case of a bouncing cosmology, there is no
trans-Planckian problem for fluctuations - their
wavelength never gets close to the Planck scale.
And like in the case of a bouncing cosmology, the
physics driving the background dynamics is
robust against our ignorance of what solves
the cosmological constant problem. These are
two advantages of the emergent scenario compared
to inflation. 

On the negative side, the origin of the large
size and entropy of our universe remains a mystery
in emergent cosmology. Also, the physics yielding
the emergent phase is not well understood in terms of
an effective field theory setting, in
contrast to the physics yielding inflation.

Whereas there are a lot of toy models for a
bouncing cosmology, there are not many
models that realize an emergent universe.
The ``String Gas Cosmology'' model discussed below
is a concrete proposal. Another recent proposal
is in the context of Galileon cosmology \cite{Nicolis}
(see \cite{Laurence} for a discussion of the
termination of the emergent phase in the context
of the model of \cite{Nicolis}). There is
also a relationship with the work of
\cite{Rubakov}.
The small number of concrete models, however,
does not mean that this approach is not promising.
I suspect that any non-perturbative approach to
quantum gravity which leads to an emergence of
space after some phase transition will lead to
a convincing realization of emergent cosmology.
 
\subsection{String Gas Cosmology}

String gas cosmology \cite{BV} (see also \cite{Perlt}, and see 
\cite{SGCrev}  for a review) is
a realization of the emergent cosmology paradigm 
which results from coupling a
gas of fundamental strings to a background space-time metric.
It is assumed that the spatial sections are compact.
For simplicity, we take all spatial directions to be toroidal and
denote the radius of the torus by $R$.

The guiding principle of string gas cosmology is to focus on 
symmetries and degrees of freedom which are new to
string theory (compared to point particle theories) and which will
be part of any non-perturbative string theory, and to use
them to develop a new cosmology. The symmetry we make use of is
{\bf T-duality}, and the new degrees of freedom are the {\bf string
oscillatory modes} (corresponding to fluctuations in the shape
of a string) and the {\bf string winding modes} (strings winding
the background space). Strings also have {\bf momentum modes} 
which correspond to the center of mass motion of the strings.
Point particles only have momentum modes.
 
The first key feature of string theory is that there is a limiting  
temperature for a gas of strings in
thermal equilibrium, the {\it Hagedorn temperature} \cite{Hagedorn}
$T_H$. This stems from the fact that the number of string oscillatory 
states increases exponentially with energy.
Thus, if we take a box of strings and adiabatically decrease the box
size, the temperature will never diverge. This is the first indication that
string theory has the potential to resolve the cosmological singularity
problem.

The second key feature of string theory upon which string gas cosmology
is based is {\it T-duality}. To introduce this symmetry, let us discuss the
radius dependence of the energy of the basic string states:
The energy of an oscillatory mode is independent of $R$, momentum
mode energies are quantized in units of $1/R$, i.e.
\be
E_n \, = \, n \mu \frac{{l_s}^2}{R} \, ,
\ee
where $l_s$ is the string length and $\mu$ is the mass per unit length of
a string. The winding mode energies are 
quantized in units of $R$, i.e.
\be
E_m \, = \, m \mu R \, ,
\ee
where both $n$ and $m$ are integers. Thus, a new symmetry of
the spectrum of string states emerges: Under the change
\be
R \, \rightarrow \, 1/R
\ee
in the radius of the torus (in units of  $l_s$)
the energy spectrum of string states is
invariant if winding
and momentum quantum numbers are interchanged
\be
(n, m) \, \rightarrow \, (m, n) \, .
\ee
The above symmetry is the simplest element of a larger
symmetry group, the T-duality symmetry group which in
general also mixes fluxes and geometry.
The string vertex operators are consistent with this symmetry, and
thus T-duality is a symmetry of perturbative string theory. Postulating
that T-duality extends to non-perturbative string theory leads
\cite{Pol} to the need of adding D-branes to the list of fundamental
objects in string theory. With this addition, T-duality is expected
to be a symmetry of non-perturbative string theory.
Specifically, T-duality will take a spectrum of stable Type IIA branes
and map it into a corresponding spectrum of stable Type IIB branes
with identical masses \cite{Boehm}. 

As discussed in \cite{BV}, the above T-duality symmetry leads to
an equivalence between small and large spaces, an equivalence
elaborated on further in \cite{Hotta,Osorio}.

\begin{figure} 
\includegraphics[height=6cm]{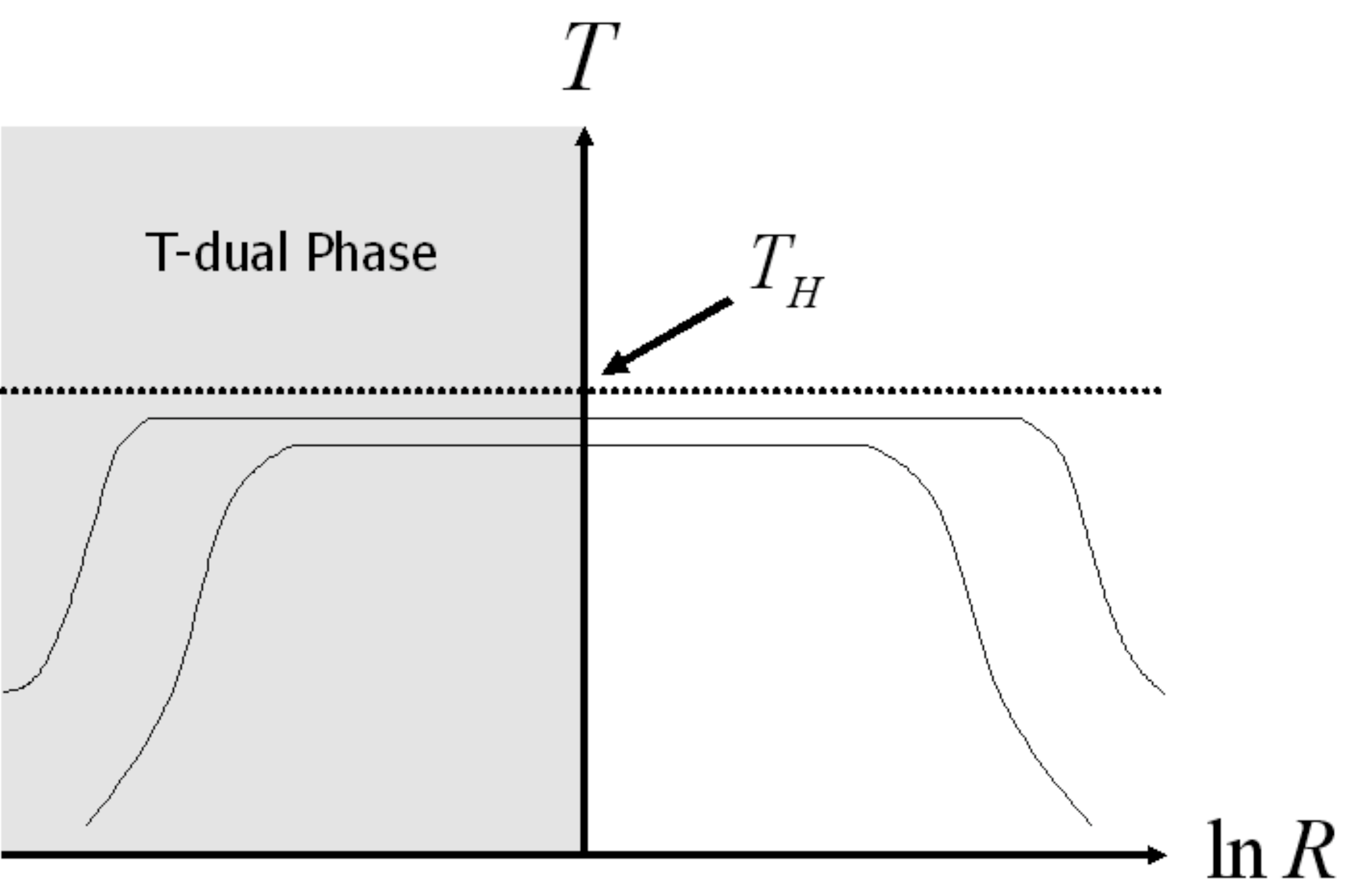}
\caption{The temperature (vertical axis) as a function of
radius (horizontal axis) of a gas of closed strings in thermal
equilibrium. Note the absence of a temperature singularity. The
range of values of $R$ for which the temperature is close to
the Hagedorn temperature $T_H$ depends on the total entropy
of the universe. The upper of the two curves corresponds to
a universe with larger entropy.}
\label{jirofig1}
\end{figure}

Let us now turn to the background cosmology which emerges from
string gas cosmology. First consider the adiabatic evolution of
a box of strings as the box radius $R$ decreases. If the initial
radius is much larger than the string length, then in
thermal equilibrium most of the energy is initially in the
momentum modes since they are the lightest ones. As
$R$ decreases, the temperature first rises as in
standard cosmology since the string states which are occupied
(the momentum modes) get heavier. However, as the temperature
approaches the Hagedorn temperature, the energy begins to
flow into the oscillatory modes and the temperature
levels off. As the radius $R$ decreases below the string scale,
the temperature begins to decrease as the energy begins to
flow into the winding modes whose energy decreases as $R$
decreases (see Figure \ref{jirofig1}). Thus, as argued in \cite{BV},  
the temperature singularity of early universe cosmology
is resolved  in string gas cosmology.

The equations that govern the background of string gas cosmology
are not known. The Einstein equations are not the correct
equations since they do not obey the T-duality symmetry of
string theory. Many early studies of string gas cosmology were
based on using the dilaton gravity equations \cite{TV,Ven,Tseytlin}. 
However, these equations are not satisfactory, either. Firstly,
we expect that string theory correction terms to the
low energy effective action of string theory become dominant
in the Hagedorn phase. Secondly, the dilaton gravity
equations yield a
rapidly changing dilaton during the Hagedorn phase (in the
string frame). Once the dilaton becomes large, it becomes
inconsistent to focus on fundamental string states rather
than brane states. In other words, using dilaton gravity as a
background for string gas cosmology does not correctly
reflect the S-duality symmetry of string theory. A
background for string gas cosmology including a rolling
tachyon was proposed \cite{Kanno1} which allows a background
in the Hagedorn phase with constant scale factor and constant
dilaton; but this construction is rather ad hoc. 
Another study of this problem was given in \cite{Sduality}.

Some conclusions about the time-temperature relation in string
gas cosmology can be derived based on thermodynamical
considerations alone. One possibility is that  $R$ starts out
much smaller than the self-dual value and increases monotonically.
From Figure \ref{jirofig1} it then follows that the time-temperature curve
will correspond to that of a bouncing cosmology. A specific
realization of this possibility in the context of a string
theory background in which the effective background equations
of motion are well justified is given in \cite{KPT}. 

Alternatively, it is possible that the universe starts out in a 
meta-stable state near the Hagedorn temperature, the {\it Hagedorn phase}, 
and then smoothly evolves into an expanding phase dominated by
radiation like in Standard Cosmology. 
Note that we are assuming that not only the scale factor but
also the dilaton is constant in time. This is the setup
which is assumed in the string gas realization of the emergent
universe scenario.

\begin{figure} 
  \includegraphics[height=2.8cm]{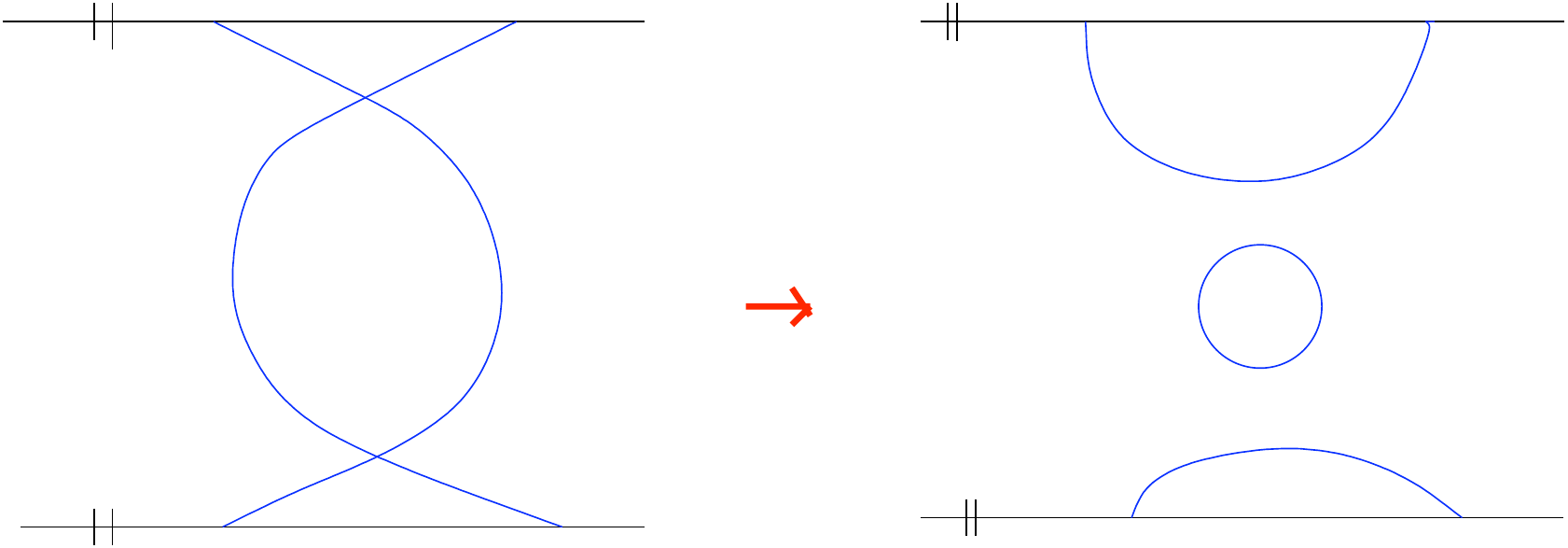}
\caption{The process by which string loops are produced via the
intersection of winding strings. The top and bottom lines
are identified and the space between these lines represents
space with one toroidal dimension un-wrapped.}
\label{decay}
\end{figure}

Note that it is the annihilation (see Figure (\ref{decay}))
of winding strings into string loops (which acts as
stringy radiation) which leads
to the transition from the early quasi-static phase to the
radiation phase of Standard Cosmology.

The evolution of the scale factor in string gas cosmology
is as in any general emergent universe scenario
(see Fig.  \ref{timeevol}). In this figure,
along the horizontal axis, the approximate equation of state
for the string gas cosmology realization of the emergent
scenario is also indicated. During the Hagedorn phase the 
pressure is negligible due to the cancellation between the 
positive pressure of the momentum
modes and the negative pressure of the winding modes, after time $t_R$
the equation of state is that of a radiation-dominated universe.

As pointed out in \cite{BV}, the annihilation process which
allows for the expansion of spatial radii
is only possible in at most three large spatial dimensions. This is
a simple dimension counting argument: string world sheets have
measure zero intersection probability in more than four large 
space-time dimensions. Hence, string gas cosmology
may provide a natural mechanism for explaining why there are
exactly three large spatial dimensions. This argument was
supported by numerical studies of string evolution in three and
four spatial dimensions \cite{Mairi} (see also \cite{Cleaver}). 
The flow of energy from
winding modes to string loops can be modelled by effective
Boltzmann equations \cite{BEK} analogous to those used to
describe the flow of energy between infinite cosmic strings and
cosmic string loops (see e.g. \cite{ShellVil,HK,RHBtoprev} for
reviews). 

There is a caveat regarding the above mechanism.
In the analysis of \cite{BEK} it was assumed that the
string interaction rates were time-independent. If the dynamics of
the Hagedorn phase is modelled by dilaton gravity, the dilaton is
rapidly changing during the phase in which the string frame scale
factor is static. As discussed in \cite{Col2,Danos} 
(see also \cite{Kabat}), in this case the
mechanism which tells us that exactly three spatial dimensions 
become macroscopic does not work. 

An important question which has to be addressed in any
model of string cosmology is what stabilizes the moduli,
in particular the sizes and shapes of the extra spatial
dimensions. In this respect string gas cosmology in the
context of heterotic string theory has some major advantages
over other approaches to string cosmology, at least in
the context of toroidal compactifications, the ones which
have been studied to date. This issue is
reviewed in detail in \cite{RHBrev5}. The basic idea
\cite{Watson} is that winding modes about the extra spatial dimensions
create an energy barrier against expansion, whereas momentum
modes cause an energy barrier against contraction. There
is hence an energetically favored value for the radius $R$ of 
an extra spatial dimension (which is typically the string length).
This is the self-dual radius. This mechanism is a special case
of the general principle of moduli trapping at enhanced
symmetry states \cite{beauty, Watson2}.

In order to avoid a cosmological constant problem, it is
important that the induced potential energy of the four-dimensional
effective field theory vanishes at the self-dual radius.
This issue has been studied in detail \cite{Subodh1, Subodh2}
in the case of heterotic superstring theory, and it was
shown that the special massless enhanced symmetry states which appear
at the self-dual radius and dominate the potential at
that point have vanishing potential energy. Thus, in
heterotic string gas cosmology the radion moduli are
dynamically stabilized. By studying the off-diagonal Einstein
equations in the presence of a string gas with both momentum
and winding modes it can also be shown \cite{Edna} that the
shape moduli are stabilized at points of extra symmetry.

The only modulus which is not stabilized by string winding and momentum modes
is the dilaton. One can \cite{Danos2} introduce 
gaugino condensation, the same mechanism used
in string inflation model building (see e.g. \cite{stringinflrev} for
a recent review) and show that this generates a 
stabilizing potential for the dilaton without interfering with the 
radion stabilization force provided
by the string winding and momentum modes. Gaugino condensation
also leads \cite{Sasmita} to supersymmetry breaking (typically at a high
energy scale).

A final comment concerns
the isotropy of the three large dimensions. In contrast to the
situation in Standard cosmology, in string gas cosmology the
anisotropy decreases in the expanding phase \cite{Watson1}.
Thus, there is a natural isotropization mechanism for the three
large spatial dimensions.

\section{Cosmological Perturbations}

\subsection{Overview}

The topic of the second lecture is the theory of cosmological
perturbations and its applications to both inflationary cosmology
and the ``un-conventional'' cosmological alternatives discussed
in the first lecture.

The theory of cosmological perturbations is the main tool of
modern cosmology. It allows us to follow the evolution of
small inhomogeneities generated in the very early universe
and propagate their evolution to the present time, which then
allows us to work out predictions of models of the early universe.
For an extensive overview of the subject the reader is referred
to \cite{MFB}, and to \cite{RHBrev} for an overview.

As we have seen in the first lecture, in many models of the very early 
Universe, in particular in inflationary cosmology, in the emergent
universe paradigm and in the matter bounce scenario, 
primordial inhomogeneities are generated in an initial phase
on sub-Hubble scales. The wavelength is then stretched
relative to the Hubble radius $H^{-1}(t)$, where $H$ is
the cosmological expansion rate, becomes larger than the Hubble
radius at some time and then propagates on super-Hubble scales until
re-entering at late cosmological times. In a majority of the current
structure formation scenarios (string gas cosmology is an exception
in this respect), fluctuations are assumed to emerge as quantum
vacuum perturbations. Hence, to describe the generation and
evolution of the inhomogeneities, both General Relativity and
quantum mechanics are required. What makes the theory of
cosmological perturbations tractable is that the amplitude
of the fractional fluctuations is small today and hence (since gravity
is a purely attractive force) that it was even smaller in the
early universe. This justifies the linear analysis of the
generation and evolution of fluctuations.

In the context of a Universe with an inflationary period, 
the quantum origin of cosmological fluctuations
was first discussed in \cite{Mukh}  and also \cite{Press,Sato} for
earlier ideas. In particular, Mukhanov \cite{Mukh} and Press 
\cite{Press} realized that
in an exponentially expanding background, the curvature fluctuations
would be scale-invariant, and Mukhanov provided a quantitative
calculation which also yielded the logarithmic deviation from
exact scale-invariance. 

Here we give a very abbreviated overview of the quantum
theory of cosmological perturbations. The reader is
referred to \cite{RHBrev} for a description which is closer
to what was presented at the Naxos school.

The basic idea of the theory of cosmological perturbations is
simple. In order to obtain the action for linearized cosmological
perturbations, we expand the action for gravity and matter to quadratic order in
the fluctuating degrees of freedom. The linear terms cancel
because the background is taken to satisfy the background
equations of motion. 

At first sight, it appears that there are ten degrees of freedom
for the metric fluctuations, in addition to the matter perturbations.
However, four of these degrees of freedom are equivalent to
space-time diffeomorphisms. To study the remaining six
degrees of freedom for metric fluctuations it proves very
useful to classify them according to how they transform under
spatial rotations. There are two scalar modes, two vector
modes and two tensor modes. At linear order in cosmological
perturbation theory, scalar, vector and tensor modes decouple.
For simple forms of matter such as scalar fields or perfect
fluids, the matter fluctuations couple only to the scalar
metric modes. These are the so-called ``cosmological
perturbations" which we study below. 

If matter has no anisotropic stress, then one of the scalar
metric degrees of freedom disappears. In addition, one
of the Einstein constraint equations couples the remaining
metric degree of freedom to matter. Thus, if there is only
one matter component (e.g. one scalar matter field), there
is only one independent scalar cosmological fluctuation mode. 

To obtain the action and equation of motion for this mode,
we begin with the Einstein-Hilbert action for gravity and the
action for matter (which we take for simplicity to be a
scalar field $\varphi$ - for the more complicated
case of general hydrodynamical fluctuations the reader is
referred to \cite{MFB})
\begin{equation} \label{action}
S \, = \,  \int d^4x \sqrt{-g} \bigl[ - {1 \over {16 \pi G}} R
+ {1 \over 2} \partial_{\mu} \varphi \partial^{\mu} \varphi - V(\varphi)
\bigr] \, ,
\end{equation}
where $R$ is the Ricci curvature scalar.

The simplest way to proceed is to work in 
longitudinal gauge, in which the metric and matter take the form
(assuming no anisotropic stress)
\begin{eqnarray} \label{long}
ds^2 \, &=& \, a^2(\eta)\bigl[(1 + 2 \phi(\eta, {\bf x}))d\eta^2
- (1 - 2 \phi(t, {\bf x})) d{\bf x}^2 \bigr] \nonumber \\
\varphi(\eta, {\bf x}) \, 
&=& \, \varphi_0(\eta) + \delta \varphi(\eta, {\bf x}) \, ,
\end{eqnarray}
where $\eta$ in conformal time. The two fluctuation variables
$\phi$ and $\delta \varphi$ must be linked by the Einstein constraint
equations since there cannot be matter fluctuations without induced
metric fluctuations. 

The two nontrivial tasks of the lengthy \cite{MFB} computation 
of the quadratic piece of the action is to find
out what combination of $\delta \varphi$ and $\phi$ gives the variable $v$
in terms of which the action has canonical kinetic term, and what the form
of the time-dependent mass is. This calculation involves inserting
the ansatz (\ref{long}) into the action (\ref{action}),
expanding the result to second order in the fluctuating fields, making
use of the background and of the constraint equations, and dropping
total derivative terms from the action. In the context of
scalar field matter, the quantum theory of cosmological
fluctuations was developed by Mukhanov \cite{Mukh2,Mukh3} and
Sasaki \cite{Sasaki}. The result is the following
contribution $S^{(2)}$ to the action quadratic in the
perturbations:
\begin{equation} \label{pertact}
S^{(2)} \, = \, {1 \over 2} \int d^4x \bigl[v'^2 - v_{,i} v_{,i} + 
{{z''} \over z} v^2 \bigr] \, ,
\end{equation}
where the canonical variable $v$ (the ``Sasaki-Mukhanov variable'' introduced
in \cite{Mukh3} - see also \cite{Lukash}) is given by
\begin{equation} \label{Mukhvar}
v \, = \, a \bigl[ \delta \varphi + {{\varphi_0^{'}} \over {\cal H}} \phi
\bigr] \, ,
\end{equation}
with ${\cal H} = a' / a$, and where
\begin{equation} \label{zvar}
z \, = \, {{a \varphi_0^{'}} \over {\cal H}} \, .
\end{equation}

As long as
the equation of state does not change over time
\begin{equation} \label{zaprop}
z(\eta) \, \sim \, a(\eta) \, .
\end{equation}
Note that the variable $v$ is related to the curvature
perturbation ${\cal R}$ in comoving coordinates introduced
in \cite{Lyth0} and closely related to the variable $\zeta$ used
in \cite{BST,BK}:
\begin{equation} \label{Rvar}
v \, = \, z {\cal R} \, .
\end{equation}

The equation of motion which follows from the action (\ref{pertact}) is
(in momentum space)
\begin{equation} \label{pertEOM2}
v_k^{''} + k^2 v_k - {{z^{''}} \over z} v_k \, = \, 0 \, ,
\end{equation}
where $v_k$ is the k'th Fourier mode of $v$. 
The mass term in the above equation is in general
given by the Hubble scale (the scale whose wave-number will
be denoted $k_H$). Thus, it immediately follows that on small
length scales, i.e. for
$k > k_H$, the solutions for $v_k$ are constant amplitude oscillations . 
These oscillations freeze out at Hubble radius crossing,
i.e. when $k = k_H$. On longer scales ($k \ll k_H$), there is
a mode of  $v_k$ which scales as $z$. This mode is the dominant
one in an expanding universe, but not in a contracting one.

Given the action (\ref{pertact}), the cosmological
perturbations can be quantized by canonical quantization (in the same
way that a scalar matter field on a fixed cosmological background
is quantized \cite{BD}). 

The final step in the quantum theory of cosmological perturbations
is to specify an initial state. Since in inflationary cosmology
all pre-existing classical fluctuations are red-shifted by the
accelerated expansion of space, one usually assumes
that the field $v$ starts out at the initial time $t_i$ mode by mode in its vacuum
state. This prescription, however, can be criticized in light of the
trans-Planckian problem for cosmological fluctuations. It assumes that 
ultraviolet modes which are continuously crossing the Planck scale cutoff
$k = m_{pl}$ are in their vacuum state, which is a strong constraint on
physics on trans-Planckian scales. 

There are two other
questions which immediately emerge: what is the initial time $t_i$,
and which of the many possible vacuum states should be chosen. It is
usually assumed that since the fluctuations only oscillate on sub-Hubble
scales, the choice of the initial time is not important, as long
as it is earlier than the time when scales of cosmological interest
today cross the Hubble radius during the inflationary phase. The
state is usually taken to be the Bunch-Davies vacuum (see e.g. \cite{BD}),
since this state is empty of particles at $t_i$ in the coordinate frame
determined by the FRW coordinates  Thus, we choose the initial
conditions
\begin{eqnarray} \label{incond}
v_k(\eta_i) \, = \, {1 \over {\sqrt{2 k}}} \\
v_k^{'}(\eta_i) \, = \, {{\sqrt{k}} \over {\sqrt{2}}} \, \, \nonumber
\end{eqnarray} 
where $\eta_i$ is the conformal time corresponding
to the physical time $t_i$.
 
Returning to the case of an expanding universe, the scaling
\begin{equation} \label{squeezing}
v_k \, \sim \, z \, \sim \, a \,
\end{equation}
implies that the wave function of the quantum variable $v_k$ which
performs quantum vacuum fluctuations on sub-Hubble scales,
stops oscillating on super-Hubble scales and instead is
squeezed (the amplitude increases in configuration space
but decreases in momentum space). This squeezing corresponds
to quantum particle production. It is also one of the two
conditions which are required for the classicalization of
the fluctuations. The second condition is decoherence which
is induced by the non-linearities in the dynamical system
which are inevitable since the Einstein action leads to
highly nonlinear equatiions (see \cite{Starob3} for an in-depth
discussion of this point, and \cite{Martineau} for related
work).

Note that the squeezing of cosmological fluctuations on
super-Hubble scales occurs in all models, in particular
in string gas cosmology and in the bouncing universe
scenario since also in these scenarios perturbations
propagate on super-Hubble scales for a long period of
time. In a contracting phase, the dominant
mode of $v_k$ on super-Hubble scales is not the one given
in (\ref{squeezing}) (which in this case is a decaying
mode), but rather the second mode which scales as $z^{-p}$
with an exponent $p$ which is positive and whose
exact value depends on the background equation of state. 

Applications of this theory in inflationary cosmology, in
the matter bounce scenario and in string gas cosmology
will be considered in the following sections.

\section{Fluctuations in Inflationary Cosmology}

We will now use the quantum theory of cosmological
perturbations developed in the previous section 
to calculate the spectrum
of curvature fluctuations in inflationary cosmology. 
The starting point are quantum vacuum initial conditions
for the canonical fluctuation variable $v_k$:
\be \label{IC}
v_k(\eta_i) \, = \, \frac{1}{\sqrt{2 k}} \, 
\ee
for all $k$ for which the wavelength is smaller than
the Hubble radius at the initial time $t_i$. 

The amplitude remains unchanged until the modes
exit the Hubble radius at the respective times
$t_H(k)$ given by
\begin{equation} \label{Hubble3}
a^{-1}(t_H(k)) k \, = \, H \, .
\end{equation}

We need to compute the power spectrum ${\cal P}_{\cal R}(k)$ 
of the curvature fluctuation ${\cal R}$ defined in (\ref{Rvar}) at
some late time $t$ when the modes are super-Hubble.
We first relate the power spectrum via the growth rate (\ref{squeezing})
of $v$ on super-Hubble scales to the power spectrum at the time 
$t_H(k)$ and then use the constancy of the amplitude
of $v$ on sub-Hubble scales to relate it to the initial conditions
(\ref{IC}). Thus
\begin{eqnarray} \label{finalspec1}
{\cal P}_{\cal R}(k, t) \, \equiv  \, k^3 {\cal R}_k^2(t) \, 
&=& \, k^3 z^{-2}(t) |v_k(t)|^2 \\
&=& \, k^3 z^{-2}(t) \bigl( {{a(t)} \over {a(t_H(k))}} \bigr)^2
|v_k(t_H(k))|^2 \nonumber \\
&=& \, k^3 z^{-2}(t_H(k)) |v_k(t_H(k))|^2 \nonumber \\
&\sim& \, k^3 \bigl( \frac{a(t)}{z(t)} \bigr)^2 a^{-2}(t_H(k)) |v_k(t_i)|^2 \, , \nonumber
\end{eqnarray}
where in the final step we have used (\ref{zaprop}) and the
constancy of the amplitude of $v$ on sub-Hubble scales. 
Making use of the condition (\ref{Hubble3}) 
for Hubble radius crossing, and of the
initial conditions (\ref{IC}), we immediately see that
\begin{equation} \label{finalspec2}
{\cal P}_{\cal R}(k, t) \, \sim \, \bigl( \frac{a(t)}{z(t)} \bigr)^2 
k^3 k^{-2} k^{-1} H^2 \, ,
\end{equation}
and that thus a scale invariant power spectrum with amplitude
proportional to $H^2$ results, in agreement with what was
argued on heuristic grounds in the overview of inflation in the
the first section. To obtain
the precise amplitude, we need to make use of the relation
between $z$ and $a$. We obtain
\be \label{amplitude}
{\cal P}_{\cal R}(k, t) \, \sim \, \frac{H^4}{\dot{\varphi}_0^2} \,
\ee
which for any given value of $k$ is to be evaluated 
at the time $t_H(k)$ (before the end of inflation). For
a scalar field potential (see following subsection)
\be
V(\varphi) \, = \, \lambda \varphi^4
\ee
the resulting amplitude in (\ref{amplitude}) is $\lambda$.
Thus, in order to obtain the observed value of the
power spectrum of the order of $10^{-10}$, 
the coupling constant $\lambda$ must
be tuned to a very small value.

\section{Matter Bounce and Structure Formation}

As we already discussed in Section 2 of these notes,
in a non-singular bouncing cosmology fluctuations
on scales relevant to current cosmological observations
have a physical wavelength which at early times during the 
contracting phase was smaller than the Hubble
radius. Hence, a causal generation mechanism for
fluctuations is possible. In fact, in \cite{Wands, Fabio1} it
was realized that fluctuations which originate on sub-Hubble
scales in their quantum vacuum state and exit the 
Hubble radius during a matter-dominated contracting
phase acquire a scale-invariant spectrum. As we review
below, this is due to the particular growth rate of
the dominant fluctuation mode in the contracting phase
which is exactly right to convert a vacuum spectrum into
a scale-invariant one. During any non-matter phase
of contraction which might follow the initial matter-dominated
phase the slope of the spectrum remains unchanged on
super-Hubble scales since all corresponding mode
functions grow by the same factor. Thus, the
spectrum of fluctuations right before the bounce is
scale-invariant. Provided that the spectrum does not
change its slope during the bounce phase, a model
falling into the matter bounce category will provide an
alternative to inflation for generating s scale-invariant
spectrum of curvature perturbations.
 
The propagation of infrared (IR) fluctuations through the 
non-singular bounce was analyzed in the case of the 
higher derivative gravity model of \cite{Biswas1}  in \cite{ABB}, 
in mirage cosmology in \cite{Saremi}, in the case of the quintom 
bounce in \cite{Cai1,LWbounce}, for a ghost condensate bounce
in \cite{Chunshan}, for a Horava-Lifshitz
bounce in \cite{HLbounce2}, and more recently \cite{BKPPT} in the
string theory bounce model of \cite{KPT}.  The result of these studies is that the
scale-invariance of the spectrum before the bounce persists
after the bounce as long as the time period of the bounce phase is
short compared to the wavelengths of the modes being considered.
Note that if the fluctuations have a thermal origin, then the condition
on the background cosmology to yield scale-invariance of the
spectrum of fluctuations is different \cite{Thermalflucts}.

\subsection{Basics}

First we will consider fluctuations in a matter bounce without
extra degrees of freedom. In this case, we need only focus
on the usual fluctuation variable $v$.
The equation of motion its Fourier mode $v_k$ is
\be \label{EOM}
v_k^{''} + \bigl( k^2 - \frac{z^{''}}{z} \bigr) v_k \, = \, 0 \, .
\ee
If the equation of state of the background is time-independent, then
$z \sim a$ and hence the negative square mass term in (\ref{EOM})
is $H^2$. Thus, on length scales smaller than the Hubble radius,
the solutions of (\ref{EOM}) are oscillating, whereas on larger
scales they are frozen in, and their amplitude
depends on the time evolution of $z$. 

In the case of an expanding universe the dominant mode of $v$ scales as
$z$. However, in a contracting universe it is the second of the
two modes which dominates. 
If the contracting phase is matter-dominated, i.e. $a(t) \sim t^{2/3}$
and $\eta(t) \sim t^{1/3}$ the dominant mode of $v$ scales as $\eta^{-1}$
and hence
\be
v_k(\eta) \, = \, c_1 \eta^2 + c_2 \eta^{-1} \, ,
\ee
where $c_1$ and $c_2$ are constants. The $c_1$ mode is the 
mode for which $\zeta$ is constant on super-Hubble scales. However,
in a contracting universe it is the $c_2$ mode which dominates and
leads to a scale-invariant spectrum \cite{Wands,Fabio1}:
\bea
P_{\zeta}(k, \eta) \, &\sim& k^3 |v_k(\eta)|^2 a^{-2}(\eta) \\
&\sim& \, k^3 |v_k(\eta_H(k))|^2 \bigl( \frac{\eta_H(k)}{\eta} \bigr)^2 \, 
\sim \, k^{3 - 1 - 2} \nonumber \\
&\sim& \, {\rm const}  \, , \nonumber 
\eea
where we have made use of the scaling of the dominant mode of $v_k$, the 
Hubble radius crossing condition $\eta_H(k) \sim k^{-1}$, and
the assumption that we have a vacuum spectrum at Hubble radius crossing.

At this point we have shown that the spectrum of fluctuations is
scale-invariant on super-Hubble scales before the bounce phase.
The evolution during the bounce depends in principle on the
specific realization of the non-singular bounce. In any concrete
model, the equations of motion can be solved 
numerically without approximation during the bounce.
Alternatively, we can solve them approximately using analytical techniques.
Key to the analytical analysis are the General Relativistic matching
conditions for fluctuations across a phase transition in the background
\cite{HV,DM}. These conditions imply that both $\Phi$ and $\tilde{\zeta}$
are conserved at the bounce, where
\be
\tilde{\zeta} \, = \, \zeta + \frac{1}{3} \frac{k^2 \Phi}{{\cal{H}}^2 - {\cal{H}}^{'}} \, .
\ee
However, as stressed in \cite{Durrer2}, these matching conditions
can only be used at a transition when the background metric
obeys the matching conditions. This is not the case if we
were to match directly between the contracting matter phase
and the expanding matter phase, as was done in early
studies \cite{Lyth, Fabio2, KOST2} of fluctuations in the Ekpyrotic scenario. 

In the case of a non-singular
bounce we have three phases: the initial contracting phase with
a fixed equation of state (e.g. $w = 0$), a bounce phase during
which the universe smoothly transits between contraction
and expansion, and finally the expanding phase with constant $w$. 
We need to apply  the matching conditions twice: 
first at the transition between the
contracting phase and the bounce phase (on both sides of
the matching surface the universe is contracting), and then between
the bouncing phase and the expanding phase. The bottom line
of the studies of  \cite{ABB, Saremi, Cai1, LWbounce, Chunshan, HLbounce2, BKPPT} 
is that on length scales large compared to the time of the bounce, the
spectrum of curvature fluctuations is not changed during the
bounce phase. Since typically the bounce time is set by
a microphysical scale whereas the wavelength of fluctuations
which we observe today is macroscopic (about $1 {\rm mm}$
if the bounce scale is set by the particle physics GUT scale),
we conclude that for scales relevant to current observations
the spectrum is unchanged during the bounce. This completes
the demonstration that a non-singular matter bounce leads
to a scale-invariant spectrum of cosmological perturbations
after the bounce provided that the initial spectrum on
sub-Hubble scales is vacuum. 

The fact that fluctuations grow both in the contracting and
expanding phase has implications for cyclic cosmologies
in four space-time dimensions: In the presence of
fluctuations, no such cyclic models are possible - the
growth of fluctuations breaks this cyclicity. As we
showed above, the spectral index of the power spectrum
of the fluctuations changes during the bounce. Hence,
four space-time-dimensional cyclic background cosmologies
are not predictive - the index of the power spectrum changes
from cycle to cycle \cite{processing}. Note that the
cyclic version of the Ekpyrotic scenario \cite{cyclic}
avoids these problems because it is not cyclic in the
above sense: it is a higher space-time-dimensional model
in which the radius of an extra dimensions evolves
cyclically, but the four-dimensional scale factor does
not.

The above analysis is applicable only as long as no new
degrees of freedom become relevant at high energy
densities, in particular during the bounce phase. In
non-singular bounce models obtained by modifying
the matter sector, new degrees of freedom
arise from the extra matter fields. They can thus
give entropy fluctuations which may compete with
the adiabatic mode studied above. In the quintom
bounce model this issue has recently been studied
in \cite{Yifu}. It was found that fluctuations in the ghost
field which yields the bounce are unimportant on large
scales since they have a blue spectrum. However,
entropy fluctuations due to extra low-mass fields
can be important. Their spectrum is also
scale-invariant, and this yields the ``matter bounce
curvaton" mechanism. 

In non-singular bouncing models obtained by modifying
the gravitational sector of the theory the identification
of potential extra degrees of freedom is more difficult.
As an example, let us mention the situation in the
case of the Horava-Lifshitz bounce. The theory has
the same number of geometric degrees of freedom
as General Relativity, but less symmetries. Thus, more
of the degrees of freedom are physical. Recall from
the discussion of the theory of cosmological perturbations
in Section 2 that there are ten total geometrical degrees
of freedom for linear cosmological perturbations, four
of them being scalar, four vector and two tensor. In Einstein
gravity the symmetry group of space-time diffeomorphisms is
generated at the level of linear fluctuations by four functions,
leaving six of the ten geometrical variables as physical -
two scalar, two vector and two tensor modes.
In the absence of anisotropic stress the number of scalar variables
is reduced by one, and the Hamiltonian constraint relates
the remaining scalar metric fluctuation to matter.

In Horava-Lifshitz gravity one loses one scalar gauge
degree of freedom, namely that of space-dependent
time reparametrizations. Thus, one expects an extra
physical degree of freedom. It has been recently
been shown \cite{Cerioni1} that in the projectable
version of the theory (in which the lapse function
$N(t)$ is constrainted to be a function of time only) the
extra degree of scalar cosmological perturbations is
either ghost-like or tachyonic, depending on parameters
in the Lagrangian. Thus, the theory appears to be
ill-behaved in the context of cosmology. However, in
the full non-projectable version (in which the lapse
$N(t, {\bf x})$ is a function of both space and time,
the extra degree of freedom is well behaved. It is
important on ultraviolet scales but decouples in the
infrared \cite{Cerioni2}.

\subsection{Specific Predictions}

Canonical single field inflation models predict very small
non-Gaussianities in the spectrum of fluctuations. One
way to characterize the non-Gaussianities is via the
three point function of the curvature fluctuation, also
called the ``bispectrum". As
realized in \cite{Xue}, the bispectrum induced
in the minimal matter bounce scenario 
(no entropy modes considered) has an amplitude which
is at the borderline of what the Planck satellite experiment
will be able to detect, and it has a special form.
These are specific predictions of the matter bounce
scenario with which the matter bounce scenario can
be distinguished from those of standard inflationary models
(see \cite{Xingang} for a recent detailed review of
non-Gaussianities in inflationary cosmology and a list
of references). In the following we give a very brief
summary of the analysis of non-Gaussianities
in the matter bounce scenario.

Non-Gaussianities are induced in any cosmological model
simply because the Einstein equations are non-linear.
In momentum space, the bispectrum contains amplitude
and shape information. The bispectrum is a function of
the three momenta. Momentum conservation implies
that the three momenta have to add up to zero. However,
this still leaves a rich shape information in the bispectrum
in addition to the information about the overall amplitude.

A formalism to compute the non-Gaussianities for the
curvature fluctuation variable $\zeta$ was developed in 
\cite{Malda}. Working in the interaction representation,
the three-point function of $\zeta$ is given to leading
order by
\begin{eqnarray}\label{tpf}
&& <\zeta(t,\vec{k}_1)\zeta(t,\vec{k}_2)\zeta(t,\vec{k}_3)>  \\
&& \,\,\,\, = \, i \int_{t_i}^{t} dt'
<[\zeta(t,\vec{k}_1)\zeta(t,\vec{k}_2)\zeta(t,\vec{k}_3),{L}_{int}(t')]>~,
\nonumber
\end{eqnarray}
where $t_i$ corresponds to the initial time before which
there are any non-Gaussianities. The square parentheses indicate
the commutator, and $L_{int}$ is the interaction Lagrangian

The interaction Lagrangian contains many terms. In particular,
there are terms containing the time derivative of $\zeta$. Each
term leads to a particular shape of the bispectrum. In an
expanding universe such as in inflationary cosmology
$\dot{\zeta} = 0$. However, in a contracting phase the time
derivative of $\zeta$ does not vanish since the dominant
mode is growing in time. Hence, there are new contributions
to the shape which have a very different form from the shape
of the terms which appear in inflationary cosmology.
The larger value of the amplitude of the bispectrum follows
again from the fact that there is a mode function which grows
in time in the contracting phase.

The three-point function can be expressed in the following
general form:
\bea
<\zeta(\vec{k}_1)\zeta(\vec{k}_2)\zeta(\vec{k}_3)> \, &=& \, 
(2\pi)^7 \delta(\sum\vec{k}_i) \frac{P_{\zeta}^2}{\prod k_i^3} \nonumber \\
&& \times {\cal A}(\vec{k}_1,\vec{k}_2,\vec{k}_3)~,
\eea
where $k_i=|\vec{k}_i|$ and ${\cal A}$ is the shape function. In
this expression we have factored out the dependence on
the power spectrum $P_{\zeta}$. In inflationary cosmology
it has become usual to express the bispectrum in terms
of a non-Gaussianity parameter $f_{NL}$. However, 
this is only useful if the shape of the three point function is
known. As a generalization, we here use \cite{Xue}
\be
{\cal |B|}_{NL}(\vec{k}_1,\vec{k}_2,\vec{k}_3) \, = \, \frac{10}{3}\frac{{\cal
A}(\vec{k}_1,\vec{k}_2,\vec{k}_3)}{\sum_ik_i^3}~.
\ee

The computation of the bispectrum is tedious. In the
case of the matter bounce (no entropy fluctuations)
the result is
\begin{eqnarray} \label{result}
{\cal A} \, &=& \, \frac{3}{256\prod k_i^2} \bigg\{ 3\sum k_i^9 +
\sum_{i\neq j}k_i^7k_j^2 \nonumber \\
&& - 9 \sum_{i\neq j}k_i^6k_j^3 +5\sum_{i \neq j}k_i^5k_j^4 \\
&& -66 \sum_{i\neq j\neq k}k_i^5k_j^2k_k^2
+9\sum_{i\neq j\neq k}k_i^4k_j^3k_k^2 \bigg\}~. \nonumber
\end{eqnarray}
This equation describes the shape which is predicted.
Some of the terms (such as the last two) are the same
as those which occur in single field slow-roll inflation,
but the others are new. Note, in particular, that
the new terms are not negligible. 

If we project the
resulting shape function ${\cal A}$ onto some
popular shape masks we 
\begin{eqnarray}
{\cal |B|}_{NL}^{\rm local} \, = \, -\frac{35}{8}~,
\end{eqnarray}
for the local shape ($k_1 \ll k_2 = k_3$). This
is negative and of order $O(1)$.
For the equilateral form
($k_1= k_2= k_3$) the result is
\begin{eqnarray}
{\cal |B|}_{NL}^{\rm equil} \, = \, -\frac{255}{64}~\, ,
\end{eqnarray}
and for the folded form ($k_1=2k_2=2k_3$) one
obtains the value
\begin{eqnarray}
{\cal |B|}_{NL}^{\rm folded} \, = \, -\frac{9}{4}~.
\end{eqnarray}
These amplitudes are close to what the
Planck CMB satellite experiment will be
able to detect.

The matter bounce scenario also predicts a change in the
slope of the primordial power spectrum on small
scales \cite{LiHong}: scales which exit the Hubble
radius in the radiation phase retain a blue
spectrum since the squeezing rate on scales larger
than the Hubble radius is insufficient to give longer
wavelength modes a sufficient boost relative to
the shorter wavelength ones.

\section{String Gas Cosmology and Structure Formation}

In this section we discuss cosmological fluctuations in one
particular realization of the emergent universe scenario,
namely string gas cosmology. In contrast to the case
of exponential inflation and the matter bounce, where
a scale-invariant spectrum emerges from initial
quantum vacuum fluctuations independent of the
specific realization of the background cosmology,
in the case of the emergent universe scenario a
scale-invariant spectrum is generated only in
the string gas cosmology realization, and in other
realizations which share some general properties
which will be mentioned at the end of this section.

\subsection{Overview}

\begin{figure}  
 \includegraphics[height=.5\textheight]{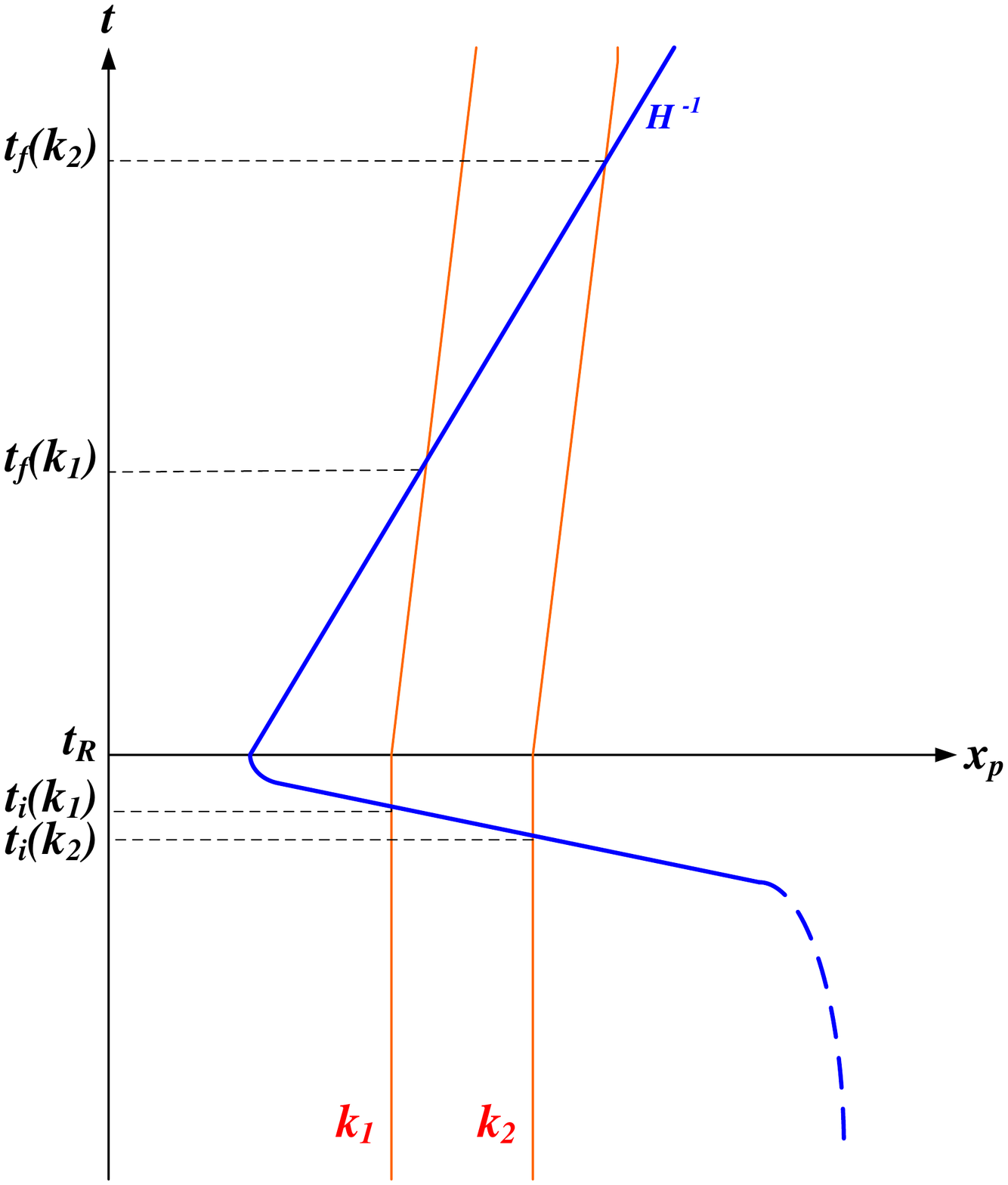}
\caption{Space-time diagram (sketch) showing the evolution of fixed 
co-moving scales in string gas cosmology. The vertical axis is time, 
the horizontal axis is physical distance.  
The solid curve represents the Einstein frame Hubble radius 
$H^{-1}$ which shrinks abruptly to a micro-physical scale at $t_R$ and then 
increases linearly in time for $t > t_R$. Fixed co-moving scales (the 
dotted lines labeled by $k_1$ and $k_2$) which are currently probed 
in cosmological observations have wavelengths which are smaller than 
the Hubble radius long before $t_R$. They exit the Hubble 
radius at times $t_i(k)$ just prior to $t_R$, and propagate with a 
wavelength larger than the Hubble radius until they re-enter the 
Hubble radius at times $t_f(k)$.}
\label{spacetimenew2}
\end{figure}

The analysis of cosmological perturbations in string gas cosmology 
(pioneered in \cite{NBV}) is based on the cosmological background 
of string gas cosmology represented in Figure \ref{timeevol}.  In turn,
this background yields the space-time diagram sketched in 
Figure \ref{spacetimenew2}. As in Figure \ref{infl1}, the vertical axis is 
time and the horizontal axis denotes the
physical distance. For times $t < t_R$, 
we are in the static Hagedorn phase and the Hubble radius is
infinite. For $t > t_R$, the Einstein frame 
Hubble radius is expanding as in standard cosmology. The time
$t_R$ is when the string winding modes begin to decay into
string loops, and the scale factor starts to increase, leading to the
transition to the radiation phase of standard cosmology. 

Let us now compare the evolution of the physical 
wavelength corresponding to a fixed co-moving scale  
with that of the Einstein frame Hubble radius $H^{-1}(t)$.
The evolution of scales in string gas cosmology is identical
to the evolution in standard and in inflationary cosmology
for $t > t_R$. If we follow the physical wavelength of the
co-moving scale which corresponds to the current Hubble
radius back to the time $t_R$, then - taking the Hagedorn
temperature to be of the order $10^{16}$ GeV - we obtain
a length of about 1 mm. Compared to the string scale and the
Planck scale, this is in the far infrared.

The physical wavelength is constant in the Hagedorn phase
since space is static. But, as we enter the Hagedorn
phase going back in time, the Hubble radius diverges to
infinity. Hence, as in the case of inflationary cosmology,
fluctuation modes begin sub-Hubble during the Hagedorn
phase, and thus a causal generation mechanism
for fluctuations is possible.

However, the physics of the generation mechanism is
very different. In the case of inflationary cosmology,
fluctuations are assumed to start as quantum vacuum
perturbations because classical inhomogeneities are
red-shifting. In contrast, in the Hagedorn phase of string gas
cosmology there is no red-shifting of classical matter.
Hence, it is the fluctuations in the classical matter which
dominate. Since classical matter is a string gas, the
dominant fluctuations are string thermodynamic fluctuations. 

Our proposal for string gas structure formation is the following 
\cite{NBV} (see \cite{BNPV2} for a more detailed description).
For a fixed co-moving scale with wavenumber $k$ we compute the matter
fluctuations while the scale in sub-Hubble (and therefore gravitational
effects are sub-dominant). When the scale exits the Hubble radius
at time $t_i(k)$ we use the gravitational constraint equations to
determine the induced metric fluctuations, which are then propagated
to late times using the usual equations of gravitational perturbation
theory. Since the scales we are interested
in are in the far infrared, we use the Einstein constraint equations for
fluctuations.

\subsection{Spectrum of Cosmological Fluctuations}

We write the metric including cosmological perturbations
(scalar metric fluctuations) and gravitational waves in the
following form (using conformal time $\eta$) 
\be \label{pertmetric}
d s^2 \, = \, a^2(\eta) \left\{(1 + 2 \Phi)d\eta^2 - [(1 - 
2 \Phi)\delta_{ij} + h_{ij}]d x^i d x^j\right\} \, . 
\ee 
As in previous sections, we are working in the longitudinal gauge
for the scalar metric perturbations and we have taken 
matter to be free of anisotropic stress. The spatial
tensor $h_{ij}({\bf x}, t)$ is transverse and traceless and represents the
gravitational waves. 

Note that in contrast to the case of slow-roll inflation, scalar metric
fluctuations and gravitational waves are generated by matter
at the same order in cosmological perturbation theory. This could
lead to the expectation that the amplitude of gravitational waves
in string gas cosmology should be generically larger than in inflationary
cosmology. This expectation, however, is not realized \cite{BNPV1}
since there is a different mechanism which suppresses the gravitational
waves relative to the density perturbations (namely the fact
that the gravitational wave amplitude is set by the amplitude of
the pressure, and the pressure is suppressed relative to the
energy density in the Hagedorn phase).

Assuming that the fluctuations are described by the perturbed Einstein
equations (they are {\it not} if the dilaton is not fixed 
\cite{Betal,KKLM}), then the spectra of cosmological perturbations
$\Phi$ and gravitational waves $h$ are given by the energy-momentum 
fluctuations in the following way \cite{BNPV2}
\be \label{scalarexp} 
\langle|\Phi(k)|^2\rangle \, = \, 16 \pi^2 G^2 
k^{-4} \langle\delta T^0{}_0(k) \delta T^0{}_0(k)\rangle \, , 
\ee 
where the pointed brackets indicate expectation values, and 
\be 
\label{tensorexp} \langle|h(k)|^2\rangle \, = \, 16 \pi^2 G^2 
k^{-4} \langle\delta T^i{}_j(k) \delta T^i{}_j(k)\rangle \,, 
\ee 
where on the right hand side of (\ref{tensorexp}) we mean the 
average over the correlation functions with $i \neq j$, and
$h$ is the amplitude of the gravitational waves \footnote{The
gravitational wave tensor $h_{i j}$ can be written as the
amplitude $h$ multiplied by a constant polarization tensor.}.
 
Let us now use (\ref{scalarexp}) to determine the spectrum of
scalar metric fluctuations. We first calculate the 
root mean square energy density fluctuations in a sphere of
radius $R = k^{-1}$. For a system in thermal equilibrium they 
are given by the specific heat capacity $C_V$ via 
\be \label{cor1b}
\langle \delta\rho^2 \rangle \,  = \,  \frac{T^2}{R^6} C_V \, . 
\ee 
The specific  heat of a gas of closed strings
on a torus of radius $R$ can be derived from the partition
function of a gas of closed strings. This computation was
carried out in \cite{Deo} (see also \cite{Ali}) with the result
\be \label{specheat2b} 
C_V  \, \approx \, 2 \frac{R^2/\ell^3}{T \left(1 - T/T_H\right)}\, . 
\ee 
The specific heat capacity scales holographically with the size
of the box. This result follows rigorously from evaluating the
string partition function in the Hagedorn phase. The result, however,
can also be understood heuristically: in the Hagedorn phase the
string winding modes are crucial. These modes look like point
particles in one less spatial dimension. Hence, we expect the
specific heat capacity to scale like in the case of point particles
in one less dimension of space \footnote{We emphasize that it was
important for us to have compact spatial dimensions in order to
obtain the winding modes which are crucial to get the holographic
scaling of the thermodynamic quantities.}.

With these results, the power spectrum $P(k)$ of scalar metric fluctuations can
be evaluated as follows
\bea \label{power2} 
P_{\Phi}(k) \, & \equiv & \, {1 \over {2 \pi^2}} k^3 |\Phi(k)|^2 \\
&=& \, 8 G^2 k^{-1} <|\delta \rho(k)|^2> \, . \nonumber \\
&=& \, 8 G^2 k^2 <(\delta M)^2>_R \nonumber \\ 
               &=& \, 8 G^2 k^{-4} <(\delta \rho)^2>_R \nonumber \\
&=& \, 8 G^2 {T \over {\ell_s^3}} {1 \over {1 - T/T_H}} 
\, , \nonumber 
\eea 
where in the first step we have used (\ref{scalarexp}) to replace the 
expectation value of $|\Phi(k)|^2$ in terms of the correlation function 
of the energy density, and in the second step we have made the 
transition to position space. 

The first conclusion from the result (\ref{power2}) is that the spectrum
is approximately scale-invariant ($P(k)$ is independent of $k$). It is
the `holographic' scaling $C_V(R) \sim R^2$ which is responsible for the
overall scale-invariance of the spectrum of cosmological perturbations.
However, there is a small $k$ dependence which comes from the fact
that in the above equation for a scale $k$ 
the temperature $T$ is to be evaluated at the
time $t_i(k)$. Thus, the factor $(1 - T/T_H)$ in the 
denominator is responsible 
for giving the spectrum a slight dependence on $k$. Since
the temperature slightly decreases as time increases at the
end of the Hagedorn phase, shorter wavelengths for which
$t_i(k)$ occurs later obtain a smaller amplitude. Thus, the
spectrum has a slight red tilt.

\subsection{Key Prediction of String Gas Cosmology}

As discovered in \cite{BNPV1}, the spectrum of gravitational
waves is also nearly scale invariant. However, in the expression
for the spectrum of gravitational waves the factor $(1 - T/T_H)$
appears in the numerator, thus leading to a slight blue tilt in
the spectrum. This is a prediction with which the cosmological
effects of string gas cosmology can be distinguished from those
of inflationary cosmology, where quite generically a slight red
tilt for gravitational waves results. The physical reason for the
blue tilt in string gas cosmology is that
large scales exit the Hubble radius earlier when the pressure
and hence also the off-diagonal spatial components of $T_{\mu \nu}$
are closer to zero.

Let us analyze this issue in a bit more detail and
estimate the dimensionless power spectrum of gravitational waves.
First, we make some general comments. In slow-roll inflation, to
leading order in perturbation theory matter fluctuations do not
couple to tensor modes. This is due to the fact that the matter
background field is slowly evolving in time and the leading order
gravitational fluctuations are linear in the matter fluctuations. In
our case, the background is not evolving (at least at the level of
our computations), and hence the dominant metric fluctuations are
quadratic in the matter field fluctuations. At this level, matter
fluctuations induce both scalar and tensor metric fluctuations.
Based on this consideration we might expect that in our string gas
cosmology scenario, the ratio of tensor to scalar metric
fluctuations will be larger than in simple slow-roll inflationary
models. However, since the amplitude $h$ of the gravitational
waves is proportional to the pressure, and the pressure is suppressed
in the Hagedorn phase, the amplitude of the gravitational waves
will also be small in string gas cosmology.

The method for calculating the spectrum of gravitational waves
is similar to the procedure outlined in the last section
for scalar metric fluctuations. For a mode with fixed co-moving
wavenumber $k$, we compute the correlation function of the
off-diagonal spatial elements of the string gas energy-momentum
tensor at the time $t_i(k)$ when the mode exits the Hubble radius
and use (\ref{tensorexp}) to infer the amplitude of the power
spectrum of gravitational waves at that time. We then
follow the evolution of the gravitational wave power spectrum
on super-Hubble scales for $t > t_i(k)$ using the equations
of general relativistic perturbation theory.

The power spectrum of the tensor modes is given by (\ref{tensorexp}):
\be \label{tpower1}
P_h(k) \, = \, 16 \pi^2 G^2 k^{-4} k^3
\langle\delta T^i{}_j(k) \delta T^i{}_j(k)\rangle
\ee
for $i \neq j$. Note that the $k^3$ factor multiplying the momentum
space correlation function of $T^i{}_j$ gives the position space
correlation function $C^i{}_j{}^i{}_j(R)$ , namely the root mean 
square fluctuation of $T^i{}_j$ in a region of radius $R = k^{-1}$. 
Thus,
\be \label{tpower2}
P_h(k) \, = \, 16 \pi^2 G^2 k^{-4} C^i{}_j{}^i{}_j(R) \, .
\ee
The correlation function $C^i{}_j{}^i{}_j$ on the right hand side
of the above equation follows from the thermal closed string
partition function and was computed explicitly in
\cite{Ali} (see also \cite{BNPV2}). We obtain
\be \label{tpower3}
P_h(k) \, \sim \, 16 \pi^2 G^2 {T \over {l_s^3}}
(1 - T/T_H) \ln^2{\left[\frac{1}{l_s^2 k^2}(1 -
T/T_H)\right]}\, ,
\ee
which, for temperatures close to the Hagedorn value reduces to
\be \label{tresult}
P_h(k) \, \sim \,
\left(\frac{l_{Pl}}{l_s}\right)^4 (1 -
T/T_H)\ln^2{\left[\frac{1}{l_s^2 k^2}(1 - T/T_H)\right]} \, .
\ee
This shows that the spectrum of tensor modes is - to a first
approximation, namely neglecting the logarithmic factor and
neglecting the $k$-dependence of $T(t_i(k))$ - scale-invariant. 

On super-Hubble scales, the amplitude $h$ of the gravitational waves
is constant. The wave oscillations freeze out when the wavelength
of the mode crosses the Hubble radius. As in the case of scalar metric
fluctuations, the waves are squeezed. Whereas the wave amplitude remains
constant, its time derivative decreases. Another way to see this
squeezing is to change variables to 
\be
\psi(\eta, {\bf x}) \, = \, a(\eta) h(\eta, {\bf x})
\ee
in terms of which the action has canonical kinetic term. The action
in terms of $\psi$ becomes
\be
S \, = \, {1 \over 2} \int d^4x \left( {\psi^{\prime}}^2 -
\psi_{,i}\psi_{,i} + {{a^{\prime \prime}} \over a} \psi^2 \right)
\ee
from which it immediately follows that on super-Hubble scales
$\psi \sim a$. This is the squeezing of gravitational 
waves \cite{Grishchuk}.

Since there is no $k$-dependence in the squeezing factor, the
scale-invariance of the spectrum at the end of the Hagedorn phase
will lead to a scale-invariance of the spectrum at late times.

Note that in the case of string gas cosmology, the squeezing
factor $z(\eta)$ for scalar metric fluctuations
does not differ substantially from the
squeezing factor $a(\eta)$ for gravitational waves. In the
case of inflationary cosmology, $z(\eta)$ and $a(\eta)$
differ greatly during reheating, leading to a much larger
squeezing for scalar metric fluctuations, and hence to a
suppressed tensor to scalar ratio of fluctuations. In the
case of string gas cosmology, there is no difference in
squeezing between the scalar and the tensor modes. 

Let us return to the discussion of the spectrum of gravitational
waves. The result for the power spectrum is given in
(\ref{tresult}), and we mentioned that to a first approximation this
corresponds to a scale-invariant spectrum. As realized in
\cite{BNPV1}, the logarithmic term and the $k$-dependence of
$T(t_i(k))$ both lead to a small blue-tilt of the spectrum. This
feature is characteristic of our scenario and cannot be reproduced
in inflationary models. In inflationary models, the amplitude of
the gravitational waves is set by the Hubble constant $H$. The
Hubble constant cannot increase during inflation, and hence no
blue tilt of the gravitational wave spectrum is possible.

A heuristic way of understanding the origin of the slight blue tilt
in the spectrum of tensor modes
is as follows. The closer we get to the Hagedorn temperature, the
more the thermal bath is dominated by long string states, and thus
the smaller the pressure will be compared to the pressure of a pure
radiation bath. Since the pressure terms (strictly speaking the
anisotropic pressure terms) in the energy-momentum tensor are
responsible for the tensor modes, we conclude that the smaller the
value of the wavenumber $k$ (and thus the higher the temperature
$T(t_i(k))$ when the mode exits the Hubble radius, the lower the
amplitude of the tensor modes. In contrast, the scalar modes are
determined by the energy density, which increases at $T(t_i(k))$ as
$k$ decreases, leading to a slight red tilt.

The reader may ask about the predictions of string gas cosmology
for non-Gaussianities. The answer is \cite{SGNG} that the
non-Gaussianities from the thermal string gas perturbations
are Poisson-suppressed on scales larger than the thermal
wavelength in the Hagedorn phase. However, if the spatial
sections are initially large, then it is possible that a network
of cosmic superstrings \cite{Witten} will be left behind. These
strings - if stable - would achieve a scaling solution (constant
number of strings crossing each Hubble volume at each
time \cite{ShellVil,HK,RHBtoprev}). Such strings give rise
to linear discontinuities in the CMB temperature maps \cite{KS},
lines which can be searched for using edge detection
algorithms such as the Canny algorithm (see \cite{Amsel}
for recent feasibility studies).

\subsection{Comments}

At the outset of this section we mentioned that not all emergent
universe scenarios will produce a scale-invariant spectrum.
For example, string gas cosmology in a non-compact
three-dimensional space will not have the holographic scaling
of the specific heat capacity and hence will not yield a
scale-invariant spectrum.

Under which conditions does our above analysis generalize?
Three conditions appear to be necessary in order to obtain
scale-invariant cosmological fluctuations from an emergent
background. Firstly, the background cosmology should have
a quasi-static early phase followed after a short transition period
by the radiation phase of Standard Big Bang cosmology.
Secondly, the evolution of cosmological fluctuations
on the infrared scales relevant to current cosmological
observations should be describable in terms of perturbed
Einstein gravity, i.e. using the formalism discussed in Section 4,
even if the background cosmology cannot. Finally, the
specific heat capacity $C_V(R)$ in a region of radius $R$
should scale holographically, i.e.
\be
C_V(R) \, \sim \, R^2 \, .
\ee

\section{Conclusions}

In these lectures I have given an overview of the matter bounce
and emergent universe scenarios of primordial cosmology. Both yield 
causal mechanisms for the generation of a scale-invariant
spectrum of cosmological perturbations, the same kind of
spectrum which is predicted by inflationary cosmology. In all
three scenarios, the fluctuations are to a good approximation
Gaussian. Thus, current cosmological observations cannot tell
these models apart. 

I have discussed specific predictions for future observations with
which the three early universe can be teased apart observationally.
The string gas cosmology realization of the emergent Universe
predicts a small blue tilt in the spectrum of gravitational waves.
Since inflationary models generically predict a red tilt, the tilt
in the gravitational wave spectrum is a very promising
characteristic. The simplest realization of the matter bounce
scenario produces a distinguished shape of the cosmological
bispectrum - and this appears to be an interesting distinctive
signal to explore.

Part of the motivation for looking for alternatives to inflation comes
from the realization that (at least current versions of) inflationary
cosmology suffers from a number of conceptual problems, in particular
a trans-Planckian problem for the fluctuations, and the
singularity problem for the background cosmology. As I
hope to have convinced the reader, these problems
are resolved both in string gas cosmology
and in the matter bounce scenario (in the latter, the
singularity problem is ``solved'' by construction).

The matter bounce and emergent universe scenarios successfully
address many of the problems of Standard Big Bang cosmology
which inflationary cosmology addresses. In particular, neither
scenario has a horizon problem. However, they do not
solve all of the problems. The biggest challenge for
the matter bounce scenario appears to be the instability to
anisotropic stress. The biggest problem of string gas cosmology
is our lack of an effective field theory which is consistent with
the cosmological background evolution which is required.
String gas cosmology also does not explain the size and
entropy of the universe.

I have focused on two alternative cosmological scenarios.
As mentioned earlier, there are more, e.g. the Ekpyrotic
universe. A goal of future research should be to find improved
realizations of all three cosmological scenarios considered
here, and also to develop new paradigms which hopefully
will have fewer conceptual problems that current ones.

\begin{acknowledgement}

I wish to thank the organizers of this school for inviting me to give
these lectures on the beautiful island of Naxos, and for their
hospitality. I wish to thank all of my collaborators on whose
work I have drawn. This work has been supported in part by funds from
an NSERC Discovery Grant and from the Canada Research Chair
program. I also acknowledge support from the Killam foundation
for the period 9/09 - 8/11.

\end{acknowledgement}
%
%
%

\end{document}